\documentstyle[12pt,draft]{article}

\newcommand{\beq}{\begin{equation}}
\newcommand{\eeq}{\end	{equation}}

\newcommand{\beqar}{\begin{eqnarray}}
\newcommand{\eeqar}{\end  {eqnarray}}

\newcommand{\benum}{\begin{enumerate}}
\newcommand{\eenum}{\end  {enumerate}}

\newcommand{\bfig}{\begin{figure}}
\newcommand{\efig}{\end  {figure}}

\newcommand{\btab}{\begin{table}}
\newcommand{\etab}{\end  {table}}

\newcommand{\AP}[1]{Ann. Phys. {\bf {#1}}}
\newcommand{\NPA}[1]{Nucl. Phys. {\bf A{#1}}}
\newcommand{\PLB}[1]{Phys. Lett. {\bf {#1}B}}
\newcommand{\PRC}[1]{Phys. Rev. {\bf C{#1}}}
\newcommand{\PRL}[1]{Phys. Rev. Lett. {\bf {#1}}}
\newcommand{\PR}[1]{Phys. Rep. {\bf {#1}}}
\newcommand{\PTP}[1]{Prog. Theor. Phys. {\bf {#1}}}

\newcommand{\bold}[1]{\mbox{\boldmath $#1$}}	

\begin{document}

\title{On the role of the $\Delta (1232)$ on the transverse\\
nuclear response in the $(e,e')$ reaction.}
\vspace{17 mm}
\author{E. Bauer\thanks{Fellow of the Consejo Nacional
de Investigaciones Cient\'{\i}ficas y T\'ecticas,
CONICET.} \\
Departamento de F\'{\i}sica, Facultad de Ciencias Exactas,\\
Universidad Nacional de La Plata,
La Plata, 1900, Argentina.}
\maketitle
\begin{abstract}
The transverse nuclear response to an electromagnetic probe which
is limited to create (or destroyed) a particle-hole ($ph$) or
delta-hole ($\Delta h$) pair is analyzed. Correlations
of the random phase approximation (RPA) type and self energy insertions
are considered. For RPA correlations we have developed a scheme
which includes explicitly the $\Delta$ and the exchange terms.
Self energy insertions over $ph$ and $\Delta h$ bubbles are
studied. Several residual interactions based on a contact
plus a ($\pi + \rho$)-meson exchange potential are used.
All calculations are performed in non-relativistic
nuclear matter. The main
effect of the $\Delta$ is to reduce the intensity over the
nuclear quasi-elastic peak. Exchange RPA terms are
very important, while the self energy depends strongly on
the residual interaction employed. We compare our final result
with data for $^{40}Ca$ at momentum transfer $q=410$
and $q=550$ MeV/c.
\end{abstract}



\newpage

\section{Introduction.}
Quasi elastic electron scattering is a powerful tool to
study the atomic nucleus. Since the experimental separation of the
inclusive longitudinal and transverse response function
\cite{kn:ba83}-\cite{kn:bl86}, a great deal of theoretical
effort was developed to understand these responses. More
recently \cite{kn:jo95}, the extraction of the experimental
points was re-analyzed. Even though, till now there
is no theoretical frame which is able to account for both
longitudinal and transverse response functions at any
momentum transfer.

Let us resume some of the theoretical efforts. Some works
assume that the nucleus is described by a Fermi
gas with a modified charge radius for individual nucleons
\cite{kn:no81}. But much of the works are based on a many body theory
\cite{kn:fu89}-\cite{kn:gi97}. The present work belongs
to this second group. Within this group some works deal
with relativistic effects \cite{kn:fu89}, the correlated
basis function \cite{kn:fa87}, meson exchange
currents (MEC) \cite{kn:al84}, \cite{kn:or81}-\cite{kn:am94},
RPA correlations \cite{kn:br87}-\cite{kn:ba97b},
Second RPA \cite{kn:co88, kn:dr90},
Extended RPA \cite{kn:ta88}-\cite{kn:ba97},
the Green function approach \cite{kn:ho80}, and the $\Delta$ degree
of freedom \cite{kn:al89}-\cite{kn:gi97}.
In fact this list is not complete, we just wanted to
mention the most relevant approaches related with the
present work. From all these references, we learn
that each effect which is considered in them, is important.
In addition, there is a strong dependence on the residual interaction
employed. The residual interaction is usually picked from
the literature, which in general corresponds to a parameterization
fixed for low energy calculations. This procedure
is questionable because an effective interaction depends on the
theory where it was adjusted \cite{kn:ba97b, kn:ce97, kn:gi97}.

That means that the search for one simple mechanism to explain data
seems to be hopeless. Many correlations like RPA, MEC, the $\Delta$
degree of freedom and so on, are all equally important.
Also, the nuclear residual interaction is unknown.
Fortunately, some simplification occurs as non relativistic
nuclear matter describe reasonable well the properties
of medium mass nuclei in the energy momentum region of
interest, once a proper Fermi momentum is used \cite{kn:am94}.

The delta play an important role in the transverse
nuclear response. In this work we have developed a method
to account for RPA correlations with the explicit inclusion
of the $\Delta$ degree of freedom
and we have also analyzed self energy insertions.
This is done for several residual interactions. As
mentioned, these contributions should be seen
as part of a set of calculations which aim should be
to reproduce both the longitudinal and the transverse responses.

The paper is organized as follows. In Section 2 we present the
formalism for RPA and self energy insertions which includes the
$\Delta$. In Section 3, we make a numerical analysis of the
different contributions. Finally, Section 4 contains the
conclusions.

\section{Formalism}
In this section we will show first the nuclear response to
an external electromagnetic probe in a general way.
Then in two sub-sections RPA correlations
and final state interactions (FSI) of the self-energy type will be
analyzed in detail.

Let us start by introducing the nuclear response function as,
\beq
R(\bold{q},\hbar \omega) = -\frac{1}{\pi} \ Im
<|{\cal O}^{\dag} G(\hbar \omega) \cal O |>
\label{eq:rn1},
\eeq
where $\bold{q}$ represents the magnitud of the
three momentum transfer by the electromagnetic probe, $\hbar \omega$
the excitation energy and $|>$ is the Hartree-Fock nuclear ground
state. Ground state
correlations beyond RPA are not analyzed in this work. The
polarization propagator is given by,
\beq
G(\hbar \omega) = \frac{1}{\hbar \omega - H + i \eta}\ -
\frac{1}{\hbar \omega + H - i \eta}\
\label{eq:gw1},
\eeq
where $H$ is the nuclear Hamiltonian. As usual, $H$ is
separated into a one-body part, $H_0$, and a residual
interaction $V$. In Eq.~(\ref{eq:rn1}) $\cal O$
represents the external probe, given by a
one body excitation operator which will
be defined soon.

We present now two projection operators $P$ and $Q$. The
action of $P$ is to project into the ground state, the one
particle-one hole ($ph$) and one delta-one hole ($\Delta h$)
configurations. While $Q$ projects into the residual
$n_p$ particle-$n_h$ hole-$n_{\Delta}$ delta configurations.
More explicitly,
\beq
P = |><| + P_N + P_{\Delta}
\label{eq:prop},
\eeq
with
\beq
P_N = \sum_{ph} |ph><ph|
\label{eq:propn},
\eeq
\beq
P_{\Delta} = \sum_{\Delta h} |\Delta h><\Delta h|
\label{eq:propd}
\eeq
and
\beq
Q  =  \sum_{ \begin{array}{c}
	     n_h \geq 2 \\ 0 \leq n_p \leq n_h
	   \end{array} }
|n_p p \; (n_h-n_p) \Delta \; n_h h><n_p p \; (n_h-n_p) \Delta \; n_h h|
\label{eq:proq},
\eeq
where we have introduced $P_N$ and $P_{\Delta}$ for convenience.
It is easy to
verify that $P + Q = 1$, $P^2 = P$,
$Q^2 = Q$, and $PQ = QP = 0$ and also,
$P_i P_j = \delta_{i j} P_i$ and $P_i Q = Q P_i = 0$ ($i = N, \Delta$).

By inserting the identity into eq.~(\ref{eq:rn1}) and noting
that the external one body operator can connect the Hartree-Fock
ground state only to the $P$ space, we have,
\beq
R_{PP}(\bold{q},\hbar \omega) = -\frac{1}{\pi} \ Im
<|{\cal O}^{\dag} \; P \; G_{PP}(\hbar \omega) \; P \; \cal O |>
\label{eq:rn2},
\eeq
where $G_{PP} \equiv P G P$. It is easy to see that,
\beq
G_{PP}(\hbar \omega) = \frac{1}{\hbar \omega - H_{PP} -
\Sigma^{PQP} + i \eta}\ -
\frac{1}{\hbar \omega + H_{PP} +
\Sigma^{PQP} - i \eta}\
\label{eq:gw2},
\eeq
where,
\beq
\Sigma^{PQP} = V_{PQ} \frac{1}{\hbar \omega -H_{QQ} + i \eta}\ V_{QP} -
	       V_{PQ} \frac{1}{\hbar \omega +H_{QQ} - i \eta}\ V_{QP}
\label{eq:self},
\eeq
with obvious definitions for $H_{PP}$, etc. As our main concern is
the effect of the $\Delta(1232)$, we
analyze the nuclear transverse response.
The matrix elements for the external operator are
then \cite{kn:so78},
\beq
<ph| \mbox{\boldmath ${\cal O} $}|> = G_E(\bold{q},\hbar \omega)
\; \frac{i}{2 m q} \;
\frac{\mu_s + \mu_v \tau_3}{2} \;
\mbox{\boldmath $q$} \times (\mbox{\boldmath $\sigma$}
\times \mbox{\boldmath $q$})
\label{eq:opn}
\eeq
and
\beq
<\Delta h| \mbox{\boldmath ${\cal O} $}|> =
G_{\Delta}(\bold{q},\hbar \omega) \;
\frac{i}{2 m q} \;
\mu_{N \Delta} T_3 \;
\mbox{\boldmath $q$} \times (\mbox{\boldmath $S$}
\times \mbox{\boldmath $q$})
\label{eq:opd}
\eeq
where $m$ is the nucleonic mass,
we have used $\mu_s=0.88$, $\mu_v=4.70$ and $\mu_{N \Delta}=3.756$.
In eq.~(\ref{eq:opn}) we have neglected the convection contribution.
In eq.~(\ref{eq:opd}) the Pauli matrices $\mbox{\boldmath $\sigma$}$
and $\tau_3$ were replaced by the corresponding transitions
matrices $\mbox{\boldmath $S$}$ and $T_3$ \cite{kn:br75}. The
electromagnetic form factors are,
\beq
G_E (\bold{q},\hbar \omega) = ( 1 +
\frac{ (\hbar c q)^2 - (\hbar \omega)^2}{(839 MeV)^2})^{-2}
\label{eq:elcfn}.
\eeq
\beq
G_{\Delta} (\bold{q},\hbar \omega) = ( 1 +
\frac{ (\hbar c q)^2 - (\hbar \omega)^2}{(1196 MeV)^2})^{-2} \;
( 1 + \frac{ (\hbar c q)^2 - (\hbar \omega)^2}{(843 MeV)^2})^{-1/2}
\label{eq:elcfd}.
\eeq
The residual interaction in the $ph$ sector is given by,
\beq
V(l) = \frac{f_{\pi NN}^2}
{\mu_{\pi}^2}
\Gamma_{\pi NN}^2 (l)
(g_{NN} \: \mbox{\boldmath $\sigma \cdot \sigma ' $} +
\tilde{g'}_{NN} (l) \mbox{\boldmath $\tau \cdot \tau ' $} \,
\mbox{\boldmath $\sigma \cdot \sigma ' $} +
\tilde{h'}_{NN} (l) \mbox{\boldmath $\tau \cdot \tau ' $} \,
\mbox{\boldmath $\sigma \cdot \widehat{l} $} \,
\mbox{\boldmath $\sigma ' \cdot \widehat{l} $} )
\label{eq:inter}
\eeq
with
\beq
\tilde{g'}_{NN} (l) = g_{NN}' -
\frac{\Gamma_{\rho NN}^2 (l)}
{\Gamma_{\pi NN}^2 (l)}\ C_{\rho NN}
\frac{l^2}{l^2 + \mu_{\rho}^2} \
\label{eq:gpt},
\eeq
\beq
\tilde{h'}_{NN} (l) =
-\frac{l^2}{l^2 + \mu_{\pi}^2} \ +
\frac{\Gamma_{\rho NN}^2 (l)}
{\Gamma_{\pi NN}^2 (l)}\ C_{\rho NN}
\frac{l^2}{l^2 + \mu_{\rho}^2} \ ,
\label{eq:hpt}
\eeq
where $\mu_{\pi} \hbar c$ ($\mu_{\rho} \hbar c$ )
is the pion (rho) rest mass and the
Landau Migdal parameters $g_{NN}$ and $g_{NN}'$ account for short
range correlations. We have used the static limit for the
interaction, where $l$ represents the
momentum transfers. For the form factor of the
$\pi NN$ ($\rho NN$ ) vertex
we have taken
\beq
\Gamma_{\pi NN, \rho NN} (l) =
\frac{\Lambda_{\pi NN, \rho NN}^2 - (\mu_{\pi, \rho} \hbar c)^2 }
{\Lambda_{\pi, \rho}^2 + (\hbar c l)^2} \ ,
\label{eq:ver}
\eeq
Numerical values for the coupling constants, masses
and form factors will be given in the next section.
Analogous expressions are obtained when deltas are involved.
In this case no Landau Migdal $g$ parameter is considered.
All the other $NN$ constants and parameters should be replace
by their corresponding $N \Delta$ and $\Delta \Delta$ values.
Also Pauli matrices must be replaced by the corresponding
transitions matrices $\mbox{\boldmath $S$}$ and $\mbox{\boldmath $T$}$
or $\mbox{\boldmath ${\cal S} $}$ and $\mbox{\boldmath ${\cal T} $}$,
the 3/2-3/2 spin matrices (see ref. \cite{kn:ce97}); depending on
the character of the mesonic vertex. Just as an example,
we consider the interaction where in one mesonic vertex
there is an incoming and outgoing $\Delta$ and in the
other vertex there is a hole. In that
case the interaction reads,

\beq
V'(l) = \frac{f_{\pi NN} f_{\pi \Delta \Delta}}
{\mu_{\pi}^2} \,
\Gamma_{\pi NN} (l)  \Gamma_{\pi \Delta \Delta} (l) \,
(\tilde{g'}_{\Delta \Delta} (l) \mbox{\boldmath $\tau \cdot {\cal T} $} \,
\mbox{\boldmath $\sigma \cdot {\cal S} $} +
\tilde{h'}_{\Delta \Delta} (l) \mbox{\boldmath $\tau \cdot {\cal T} $} \,
\mbox{\boldmath $\sigma \cdot \widehat{l} $} \,
\mbox{\boldmath ${\cal S} \cdot \widehat{l} $} )
\label{eq:inted}
\eeq

Equations (\ref{eq:rn2})-(\ref{eq:self}) generates the standard
RPA and self energy contributions. We analyze separately now
these two kind of correlations.

\subsection{RPA correlations:}
The aim of this subsection is to present a RPA formalism in
nuclear matter with the $\Delta(1232)$ and which explicitly
includes exchange terms. As a first step we neglect self
energy insertions (or equivalently we turn off the $Q$-space).
Eq.~(\ref{eq:gw2}) becomes,
\beq
G_{PP}(\hbar \omega) = \frac{1}{\hbar \omega - H_0 - V + i \eta}\ -
\frac{1}{\hbar \omega + H_0 + V - i \eta}\
\label{eq:gw3},
\eeq
where we have split the nuclear Hamiltonian. The
presence of $V$, the residual interaction,
makes $G_{PP}$ to be nondiagonal in
$P$-space. To treat this, the standard Dyson equation is employed,
\beqar
G_{PP} & = & G_{PP}^0 \; + \; G_{PP}^0 \; V \; G_{PP}
\nonumber \\
       & = & G_{PP}^0 \; + \; G_{PP}^0 \; V \; G_{PP}^0 \; + \;
		 G_{PP}^0 \; V \; G_{PP}^0 \; V \; G_{PP}^0 \; + \; ...
\label{eq:dy1},
\eeqar
where $G_{PP}^0$ results from replacing the total Hamiltonian
by its one body part. Eq.~(\ref{eq:dy1}) contains both direct and
exchange terms. If one keeps only direct terms or if one uses a
contact interaction, then Eq.~(\ref{eq:dy1}) can be
easily sum up to infinite order, leading to the ring series.
This sum can not be done when exchange terms for a finite range
interaction are included. Even this is a well know fact, let us
show it in a rather elementary way, as it will simplified the
further discussion.

We consider the firsts two terms in the second line of Eq.~(\ref{eq:dy1})
and we replace $P$ by it definition of Eq.~(\ref{eq:prop}). In
addition, let us analyzed $ph$ configurations only. Taking the matrix
elements for the firsts perturbative terms and inserting them into
Eq.~(\ref{eq:rn2}), the response function becomes,
\beqar
R_{PP} & = & -\frac{1}{\pi}\ Im \{ \sum_{ph} <|{\cal O}^{\dag} |ph>
<ph| G^0 |ph> <ph| {\cal O} |> \; + \;
\nonumber \\ [5.mm]
& &
\sum_{ph, p'h'}
<|{\cal O}^{\dag} |ph>
<ph| G^0 |ph>
<ph| V |p'h'>_{D+E} \; \times
\nonumber \\ [5.mm]
& & <p'h'| G^0 |p'h'>
<p'h'| {\cal O} |> \; + \; ... \; \}
\label{eq:rnfrpa}.
\eeqar
using momentum conservation as shown in Fig. 1, direct and exchange
matrix elements of the residual interaction can be draw as,
\beq
<(\bold{h+q}),\bold{h}| V | (\bold{h'+q'}),\bold{h'}>_D \equiv
{\cal V}_D (q)
\label{eq:vdir}
\eeq
and
\beq
<(\bold{h+q}),\bold{h}| V | (\bold{h'+q'}),\bold{h'}>_E \equiv
{\cal V}_E (|\bold{h-h'}|)
\label{eq:vexc}.
\eeq
Finally, the response function becomes,
\beqar
R_{PP} & = & -\frac{1}{\pi}\ Im \{ \sum_{ph} <|{\cal O}^{\dag} |ph>
<ph| G^0 |ph> <ph| {\cal O} |> \; + \;
\nonumber \\ [5.mm]
& &
( \sum_{ph}
<|{\cal O}^{\dag} |ph>
<ph| G^0 |ph>) \; {\cal{V}}_D (q)
\nonumber \\ [5.mm]
& &
( \sum_{p'h'}
<p'h'| G^0 |p'h'>) <p'h'|{\cal O} |> ) \; + \;
\nonumber \\ [5.mm]
& &
\sum_{ph, p'h'}
<|{\cal O}^{\dag} |ph>
<ph| G^0 |ph> \; {\cal V}_E (|\bold{h-h'}|) \;
\nonumber \\ [5.mm]
& & <p'h'| G^0 |p'h'>
<p'h'| {\cal O} |> \; + \; ... \; \}
\label{eq:rnsrpa}.
\eeqar
it is trivial to extend the procedure to higher orders or
$\Delta h$ configurations. As seen in the second term
of this equation, direct terms split into common factors. This
is not the case of the third term due to the presence of
${\cal V}_E (|\bold{h-h'}|)$, except if one uses a contact
interaction
\footnote{Also the same holds when ${\cal V}_E (|\bold{h-h'}|)$
is a separable interaction which is not the case of
a $(\pi + \rho)$-meson exchange potential.}.

Exchange terms of the RPA type happens to be important (see
refs. \cite{kn:sh89} and \cite{kn:ba96}) and as shown, they
can not be sum up to infinite order. One has to evaluate
each exchange term explicitly and in practice this can be
done up to second order. In the next section, we will show
that keeping exchange terms up to second order is not
in general a good approximation. Evidently if one
choose an arbitrarily small residual interaction a fast
convergence to the RPA series will be obtained from its
firsts perturbative terms.

Let us go back to our scheme which accounts for RPA correlations
in nuclear matter with the explicit inclusion of exchange terms.
The scheme is an extension of the one developed in ref. \cite{kn:ba96}
to include $\Delta h$ excitations and it is based on three elements.
First, it is possible to sum up to infinite order exchange terms
for a contact interaction. Second, it is possible to sum up to
infinite order direct terms for any interaction and for some
particular interactions the first two perturbative terms accounts
for the full sum. Finally, even the exchange terms of
a finite range interaction can be evaluated up to second
order, it is plausible to expect that higher order terms
will be negligible small if it is the case for theirs corresponding
direct ones and they keep smaller than them.

Thus, we divide the residual interaction
as follows,
\beq
V=V_1+V_2
\label{eq:in1},
\eeq
where $V_1$ is a contact interaction and
$V_2$ contains a contact plus the exchange of
the $(\pi+\rho)$-mesons (or any finite range interaction).
The contact term in $V_2$ is chosen to
fulfill the second and third conditions mentioned above.
An additional constrain is that the remaining contact term
($V_1$) allows a perturbative treatment.

The polarization propagator of Eq.~(\ref{eq:gw3})
can now be written as
\beq
G_{PP} \; = \; G_{1 \; PP} \; + \; G_{2 \; PP} \; + \; G_{12 \; PP}
\label{eq:gw4},
\eeq
where,
\beqar
G_{1 PP} & = & G^0_{PP} \; + \; G^0_{PP} V_1 G_{1 PP},
\label{eq:gw5a} \\ \nonumber \\
G_{2 PP} & = & G^0_{PP} V_2 G^0_{PP} \; + \;
G^0_{PP} V_2 G^0_{PP} V_2 G^0_{PP},
\label{eq:gw5b} \\ \nonumber \\
G_{12 PP} & = & G^0_{PP} V_2 G^0_{PP} V_1 G^0_{PP} \; + \;
G^0_{PP} V_2 G^0_{PP} V_1 G^0_{PP}  V_1 G^0_{PP}
\; + \; ...
\label{eq:gw5c} \\ \nonumber
\eeqar
Inserting now Eq.~(\ref{eq:gw4}) into Eq.~(\ref{eq:rn2}) one
can define three different contributions to the
response function, $R_1$, $R_2$ and $R_{12}$,
associated to $G_1$, $G_2$ and $G_{12}$, respectively.
Let us analyzed each contribution separately.

The $R_1$ contribution is simply the ring approximation (RA),
with the inclusion of the $\Delta h$ space. The solution
of Eq.~(\ref{eq:gw5a}) is given by (see Ref. \cite{kn:al89}
and \cite{kn:sh86}),
\beq
G_{1 PP} = ( I \; - \; G^0_{PP} \; V_{1 PP})^{-1} \; G^0_{PP}
\label{eq:gw5as},
\eeq
where
\beq
 G^0_{PP} = \left( \begin{array}{cc}
	 G^0_{NN} & 0 \\
	 0	  & G^0_{\Delta \Delta}
	       \end{array} \right) \makebox[.6in][l]{,}
   V_{1 \; PP} = \left( \begin{array}{cc}
	 V_{1 \; NN}	   & V_{1 \; \Delta N}	\\
	 V_{1 \; N \Delta} & V_{1 \; \Delta \Delta}
	       \end{array} \right)
\nonumber
\eeq
and
\beq
 I = \left( \begin{array}{cc}
	 P_{N} & 0 \\
	 0     & P_{\Delta}
	       \end{array} \right).
\nonumber
\eeq

We have split up the projection operator into its
components $P_N$ and $P_{\Delta}$. Finally, the contact
contribution to the response becomes,
\beq
 R_1 (\bold{q},\hbar \omega) = - \frac{1}{\pi} \ Im \;
<| \; ({{\cal O}^{\dag}}_N,
{{\cal O}^{\dag}}_{\Delta} )
\left( \begin{array}{cc}
	 G_{1 NN}	& G_{1 \Delta N}  \\
	 G_{1 N \Delta} & G_{1 \Delta \Delta}
	       \end{array} \right)
\left( \begin{array}{c}
	 {\cal O}_N	   \\
	 {\cal O}_{\Delta}
	       \end{array} \right)  \; |>
\label{eq:r1}
\eeq
where we have defined ${\cal O}_N=< ph| {\cal O} |>$
and ${\cal O}_{\Delta}=< \Delta h| {\cal O} |>$.

As mentioned $V_1$ is a contact interaction and contains
both direct and exchange contributions. How to build
this direct plus exchange interaction is described in
Appendix A. Obviously Eq.~(\ref{eq:r1}) is also valid
for direct terms of any interaction. A graphical representation
of the firsts perturbative terms stemming from this equation
is given in Fig.~2.

Also in Fig.~2 we show the $R_2$ contribution. In
Appendix B, we list analytical expressions for the main terms
contributing to $R_2$, given by the standard
rules for Golstone diagrams.

Finally, also some of the lower order
contributions to $R_{12}$ are presented in Fig.~2.
In our scheme $V_2$ is
included up to second order and $V_1$ up to infinite order.
From the three contributions, $R_{12}$ has the most complex
structure. Formally, the analysis is simplified due to the
fact that a direct plus exchange contact interaction
can not connect the $P_N$ and $P_{\Delta}$ spaces (see
Appendix A).

In Fig.~3 we show in detail the contributions to
$R_{12}$ limiting $V_2$ up to first order. It was further
split up into three contributions, $(R_{12})_{NN}$,
$(R_{12})_{N \Delta}$ and $(R_{12})_{\Delta \Delta}$;
depending on the configuration where the external operator
is attached. Each line in Fig.~3 represents a sum up to
infinite order in $V_1$. Let us call by $x$ the $ph$
bubble and by $z$ the corresponding
$\Delta h$ one. Both are defined in Appendix B.
Also we denote by $B_{1 NN}$,
$B_{1 N \Delta}$ and $B_{1 \Delta \Delta}$ ($B_{2 NNN}$
$B_{2 NN \Delta}$, $B_{2 N \Delta \Delta}$ and
$B_{2 \Delta \Delta \Delta}$) the direct plus exchange
response functions which are
first order in $V_2$ (second order in $V_2$). In
each case, subindex $N$ or $\Delta$ refers to the particular
$P$ space which builds the contribution. For instance, $B_{1 NN}$
is the sum of graphs $(B_{1 NN})_D$ plus $(B_{1 NN})_E$ of
Fig.~1. Expressions for each of these contributions are
given in Appendix B.

We show now $R_{12}$. As mentioned,
\beq
R_{12} \; = \; (R_{12})_{NN} \; + \; (R_{12})_{N \Delta}
\; + \; (R_{12})_{\Delta \Delta}
\label{eq:r12},
\eeq
where,
\beqar
(R_{12})_{NN} & = & -\frac{1}{\pi} \ Im \; \{
\frac{1}{2} \, {\cal V'}_{1 N}  \, (B_{1 NN}
+ 3 \; B_{2 NNN} ) \; \frac{x \; (2- {\cal V'}_{1 N} \; x)}
{(1-{\cal V'}_{1 N} \; x)^2} \}
\label{eq:r12n}, \\ \nonumber \\ [5. mm]
(R_{12})_{N \Delta}& = & -\frac{1}{\pi} \ Im \; \{
{\cal V'}_{1 N} \, {\cal V'}_{1 \Delta} \,
[B_{1 N \Delta}(\frac{1}{1-{\cal V'}_{1 N} \; x} \;
\frac{1}{1-{\cal V'}_{1 \Delta} \; z} - 1)
\; + \; \nonumber \\ [5. mm]
& & \; + \; B_{2 NN \Delta}
\frac{x \; (2- {\cal V'}_{1 N} \; x)}{(1-{\cal V'}_{1 N} \; x)^2} \;
\frac{1}{1-{\cal V'}_{1 \Delta} \; z}
\; + \; \nonumber \\ [5. mm]
& & \; + \; B_{2 N \Delta \Delta}
\frac{z \; (2- {\cal V'}_{1 \Delta} \; z)}
{(1-{\cal V'}_{1 \Delta} \; z)^2} \;
\frac{1}{1-{\cal V'}_{1 N} \; x} \}
\label{eq:r12nd}, \\ \nonumber \\ [5. mm]
(R_{12})_{\Delta \Delta}& = & -\frac{1}{\pi} \ Im \;
\{ \frac{1}{2} \, {\cal V'}_{1 \Delta} \,
(B_{1 \Delta \Delta}
+ 3 \; B_{2 \Delta \Delta \Delta } ) \;
\frac{z \; (2- {\cal V'}_{1 \Delta} \; z)}
{(1-{\cal V'}_{1 \Delta} \; z)^2} \}
\label{eq:r12dd} \\ \nonumber
\eeqar
and
\beqar
{\cal V'}_{1 N} & = & 8 \, \frac{m c^2}{(2 \pi)^2} \;
\frac{f_{\pi NN}^2}{4 \pi} \; \frac{1}{\mu_{\pi}^2 k_F} \;
g'_{1 \; NN} \;
\Gamma_{\pi NN}^2 (Q)
\label{eq:v1n} \\ \nonumber \\
{\cal V'}_{1 \Delta} & = & \frac{32}{9} \; \frac{m c^2}{(2 \pi)^2} \;
\frac{f_{\pi \Delta N}^2}{4 \pi} \; \frac{1}{\mu_{\pi}^2 k_F}
\; g'_{1 \; \Delta \Delta}
\Gamma_{\pi \Delta N}^2 (Q)
\label{eq:v1d} \\ \nonumber
\eeqar
where $g'_{1 \; NN}$ and $g'_{1 \; \Delta \Delta}$
are the Landau Migdal direct plus exchange terms for the
contact interaction $V_1$.
Some third order contributions ($B_{2 N \Delta N}$ in Eq.~(\ref{eq:r12n})
and $B_{2 \Delta N \Delta}$) in Eq.~(\ref{eq:r12dd})), where
neglected as they are negligible small.

We write now the RPA contribution to the response function as,
\beq
R_{PP}^{RPA} \; = \; \tilde{R}_1 \; + \; R_{12} \; + \; R_2
\label{eq:rpprpa},
\eeq
where we have redefined $\tilde{R}_1 \equiv R_1 - R_{PP}^0$,
$R_{PP}^0$ being the free response. This was done for
convenience because the free response will be included within
the self energy contribution.

Finally, if the residual interaction $V$,
fulfills the three conditions quoted at the beginning of this
sub section then $R_2$ accounts with good accuracy for
the full direct plus exchange RPA. As we will show in the next
section in general this is not the case. Then $V$ should be
split into two terms $V_1$ and $V_2$, where the first is
a contact interaction while $V_2$ is chosen
in such a way that a fast
convergence to the RPA series is achieved.
In addition to the perturbative
terms in $V_2$, it appears two new terms
$R_1$ and $R_{12}$. In this sub section we have presented
a scheme to deal with them.

\subsection{Self Energy insertions:}
We consider now self energy insertions over a single $ph$
or $\Delta h$ bubble. This means to neglect $V_{PP}$ in
Eq.~(\ref{eq:gw2}). That is,
\beq
G_{PP}(\hbar \omega) = \frac{1}{\hbar \omega - H_0 -
\Sigma^{PQP} + i \eta}\ -
\frac{1}{\hbar \omega + H_0 +
\Sigma^{PQP} - i \eta}\
\label{eq:gw5},
\eeq
we insert Eq.~(\ref{eq:gw5}) into Eq.~(\ref{eq:rn2}) and
using the definition of $P$,
we write the response function which contains the Lindhard ($L$) plus
self energy ($SE$) contributions as,
\beq
R_{PP}^{L+SE} \; = \; R_{NN}^{L_N+SE} \; + \; R_{N \Delta}^{SE}
\; + \; R_{\Delta \Delta}^{L_{\Delta}+SE}
\label{eq:rse},
\eeq
where,
\beqar
R_{NN}^{L_N+SE} & = & -\frac{1}{\pi} Im \! \! \sum_{ph,p'h'}
{{\cal O}_N}^{\dag}<ph| \frac{1}{\hbar \omega - H_0 -
\Sigma^{P_N Q P_N} + i \eta} |p'h'> {\cal O}_N
\label{eq:rsen}, \\ \nonumber \\ [5. mm]
R_{N \Delta}^{SE} & = & -\frac{1}{\pi} \ Im \! \! \sum_{ph,\Delta h}
{{\cal O}_N}^{\dag}<ph| \frac{1}{\hbar \omega - H_0 -
\Sigma^{P_N Q P_{\Delta}} + i \eta} |\Delta h>
{\cal O}_{\Delta}
\label{eq:rsed}, \\ \nonumber \\ [5. mm]
R_{\Delta \Delta}^{L_{\Delta}+SE} & = & -\frac{1}{\pi} \ Im
\! \! \sum_{\Delta h, \Delta' h'}
{{\cal O}_{\Delta}}^{\dag}<\Delta h| \frac{1}{\hbar \omega - H_0 -
\Sigma^{P_{\Delta} Q P_{\Delta}} + i \eta} |\Delta' h'>
{\cal O}_{\Delta}
\label{eq:rsend} \\ \nonumber
\eeqar
for simplicity only forward going contribution were shown. The
notation $L_N$ refers to the Lindhard function while $L_{\Delta}$
represents the $\Delta h$ bubble.

Self energy insertions are not diagonal in $P$-space. We must
consider diagonal and non diagonal self energy insertions,
shown in Figs. 4 and 5, respectively. In these figures only
second order contributions are presented. Non diagonal contributions
which connects $P_N$ and $P_{\Delta}$ spaces with
the self energy attached to the same fermionic line cancel
due to the isospin summation. We analyzed now in detail diagonal
self energy insertions. Regarding non diagonal terms, there
are 16 different contributions to it when keeping terms up
to second order. They will be evaluated in the next section
and formal expressions are obtained from the standard rules for
Goldstone diagrams. In Fig. 5 we show some of these
diagrams.

Diagonal second order self energy contributions are divergent.
To overcome this difficulty one possibility is to implement
some renormalization procedure \cite{kn:al91}.
The other alternative is to
sum self energy up to infinite order, as was done in
previous works (see refs. \cite{kn:ba95} and \cite{kn:ba97}).
We extend now the procedure developed in ref. \cite{kn:ba95}
to include the $\Delta(1232)$.

From Eq.~(\ref{eq:self}) diagonal matrix elements for the
self energy can be draw as,
\beqar
<P| \Sigma^{PQP} |P> & = &
\sum_Q
<P| V |Q>_{D + E}
\frac{1}{\hbar \omega - E_{QQ}^0 + i \eta}
<Q| V |P>_{D + E} - \nonumber \\
& - & \sum_Q
<P| V |Q>_{D + E}
\frac{1}{\hbar \omega + E_{QQ}^0 - i \eta}
<Q| V |P>_{D + E}
\label{eq:selfd},
\eeqar
where $|P>$ can be either a $ph$ or $\Delta h$ configuration, $|Q>$
is any of the $Q$-configurations and
$ H_{0} |Q> \; = \; E_{QQ}^0 |Q> $.

In Fig. 6, we have isolated a graphical representation
of the direct self energy $\Sigma^{PQP}$ restricting
ourselves to the $Q$-space. Notice that each contribution
has a closed fermionic loop (a $ph$ or $\Delta h$ bubble)
and two open fermionic lines, one of which is always
a hole. Performing the summation over spin and isospin and
making the conversion of sums over momenta to integrals it
is easy to see that these matrix elements can be written
as a function of the energy momentum of the
external probe and the momentum of the hole line,
\beq
<P| \Sigma^{PQP} |P> =
\Sigma^{PQP} (\bold{h},\bold{q},\hbar \omega)
\label{eq:defs}.
\eeq
Through the numerical analysis it turns out that
the dependence of the self energy over the
hole momentum is not very strong. This allows us to
make an average over it,
\beq
\Sigma^{PQP} (\bold{h} ,\bold{q},\hbar \omega)
\stackrel{[average \: over \bold{h}]}{\longrightarrow}
\Sigma^{PQP} (\bold{q},\hbar \omega)
\label{eq:aver}.
\eeq
as described in Appendix C. This approximation is
not so good when the self energy is attached to a hole
line, but this contribution is by itself very small.
In Appendix C we also show explicit expressions for each
self energy contributions. Let us call,
\beq
\Sigma^{NN} (\bold{Q},\nu) = \sum_{i=1}^{4}
\Sigma^{P_N Q_i P_N} (\bold{Q},\nu)
\label{eq:senn}.
\eeq
and
\beq
\Sigma^{\Delta \Delta} (\bold{Q},\nu) = \sum_{i=1}^{4}
\Sigma^{P_{\Delta} Q_i P_{\Delta}} (\bold{Q},\nu)
\label{eq:sedd}.
\eeq
where the different $Q_i$, are defined in Fig. 6.
We have used dimensionless quantities $\bold{Q=q}/k_F$
and $\nu=\hbar \omega/ 2 \varepsilon_F$;
$k_F$ and $\varepsilon_F$ being the Fermi
momentum and energy, respectively.

We defined now the functions $\Lambda(\bold{Q},\nu)$
and $\Gamma(\bold{Q},\nu)$ for the real and imaginary part
of the self energy,
\beq
Re \; \Sigma^{NN (\Delta \Delta)} (\bold{Q},\nu) \equiv
\Lambda^{NN (\Delta \Delta)} (\bold{Q},\nu)
\label{eq:sedlr}
\eeq
and
\beq
Im \; \Sigma^{NN (\Delta \Delta)} (\bold{Q},\nu) \equiv
- \frac{1}{2}
\Gamma^{NN (\Delta \Delta)} (\bold{Q},\nu)
\label{eq:sedlr3}.
\eeq
The diagonal contributions from Eq.~(\ref{eq:rsen})
and Eq.~(\ref{eq:rsend}) gives,
\beqar
(R_{PP}^{L_P+SE})^{diag.} & = & -\frac{1}{\pi} Im
\sum_P <|{\cal O}^{\dag}|P>
\frac{(2 \varepsilon_F)^{-1}}
{\nu - \varepsilon_P - \Lambda^{PP}(\bold{Q},\nu)
+  i \: \Gamma^{PP}(\bold{Q},\nu)/2 } \nonumber \\
& & \times <P|{\cal O}|>
\label{eq:selfdi}.
\eeqar
where $ H_{0} |P> \; = \; 2 \varepsilon_F \; \varepsilon_P \; |P> $.
This expression can be rewrite as (see ref. \cite{kn:ba95} for details),
\beq
(R_{PP}^{L_P+SE})^{diag.} = \int_{0}^{\infty} dE
R_{PP}^0 (\bold{Q}, E) \frac{1}{2 \pi} \
\frac{\Gamma^{PP} (\bold{Q},\nu)}{(E - \nu +
\Lambda^{PP} (\bold{Q},\nu))^2 +
(\Gamma^{PP}(\bold{Q},\nu)/2 )^2} \ \
\label{eq:selfi}.
\eeq
where the free response is,
\beq
R_{PP}^0 (\bold{Q}, E) = (2 \varepsilon_F)^{-1}
\sum_P
|<P| {\cal O} |>|^2
\delta(\varepsilon_P - E)
\label{eq:rpp0},
\eeq

Finally, two points deserve special attention. The
first one refers to the exchange terms to the self energy.
Diagonal exchange contributions are shown in Fig. 7 for $Q_1$ space and
explicit expressions can be found in Appendix C. They were included
just for completeness but its contribution is very
small. Non diagonal exchange contributions where analyzed
in ref. \cite{kn:ba97} for the $ph$ sector. In the next
section we will see that direct non diagonal contributions
are in itself small. For this reason, these exchange
contributions were not considered at all.

As a second point we want to consider again Eq.~(\ref{eq:rn2}).
In the first subsection we have neglected the self
energy and we have solved the problem of a general
Hamiltonian leading to the RPA correlations.
In the second subsection we have dressed $ph$ and $\Delta h$-
bubbles with self energy insertions.
In ref. \cite{kn:ba95} we have used these dressed bubbles
to re-calculate the ring series. We will not attempt to do this
kind of calculations here due to the presence of  exchange
terms in the RPA series which will give rice to some
exchange terms of third and higher order very difficult
to evaluate. A scheme which accounts for these contributions
is not available at present.

From both subsections, the response function is expressed as
the sum of Eqs.~(\ref{eq:rpprpa}) and (\ref{eq:rse}), as
\beq
R_{PP} \; = \; R_{PP}^{RPA} \; + \; R_{PP}^{L \, + \, SE}
\label{eq:rpptot}.
\eeq
In the next section we will analyzed numerically these contributions.

\section{Results and Discussion.}

In this section we give numerical values and discuss the different
terms coming from our scheme and in particular we analyze the influence
of the $\Delta (1232)$. We will follow the notation
and the ordering of the last section.
At the end of this section we compare our results with
data from the transverse response of $^{40}Ca$.
All calculations were done for nuclear matter with an effective
Fermi momentum $k_F=235 \; MeV/c$ \cite{kn:am94}.

For the parameters entering into our theory we have set at
140 MeV (770 MeV) the mass of the pion (rho meson). The
pion coupling constant $f_{\pi NN}^2/4 \pi$=0.081,
$f_{\pi \Delta N}$=2 $f_{\pi NN}$ and
$f_{\pi \Delta \Delta}$= $\frac{4}{5} \, f_{\pi NN}$.
For the rho meson we have used
$C_{\rho NN} = C_{\rho \Delta N} = C_{\rho \Delta \Delta}$ = 2.18.
The different mesons cut-offs at the vertices are set to
$\Lambda_{\pi NN}$ = 1300 MeV/c, $\Lambda_{\rho NN}$ = 1750 MeV/c
while all the remaining cut-offs are set to 1000 MeV/c.
For the Landau Migdal parameters we have taken
$g_{NN}=0.3$ and $g'_{NN}=0.7$, while
different values for $g'_{\Delta N} $ and
$g'_{\Delta \Delta}$ are considered.

As mentioned, we discuss now in two subsections the RPA
correlations and the self energy insertions, respectively.
At the end of the second subsection, we
consider the RPA plus self energy results. All this is done
for three energy regions: the quasi elastic peak, the
$\Delta(1232)$-peak and the 'dip' region (that is, the
region in between the two previous ones). We name the
first and second regions as the $NN$ and $\Delta \Delta$ sectors.

It is more convenient
to analyze the structure function rather than the response
function which means to put the electromagnetic form factors
equal to one
($G_E=G_{\Delta}$=1). This allows a better understanding
of the different contributions entering into our scheme.
Even thought, in the last section we have preferred to
show the response function. As there is two different electromagnetic form
factors, it could be confuse to construct the response
function from the structure function. That is, some terms
should be multiplied by $G_E^2$, others by $G_{\Delta}^2$ and
finally others by $G_E \, G_{\Delta}$.
In what follows, we analyzed then
the structure function per unit volume except when we compare
with data.

\subsection{RPA results:}
We start by analyzing the validity of the scheme presented in
subsection 1.1. The transverse structure function per unit
volume at momentum transfer $q=410 \, MeV/c$ is studied in
detail. The first step is to build a contact plus finite range
interaction which fulfills all the conditions required for
$V_2$. As the finite range piece of the interaction is fixed
by the $(\pi + \rho)$-meson exchange potential, the problem
is reduced to find an appropriate set of Landau Migdal parameters
(which represent the contact piece of the interaction).
We define,
\beq
g'_{PP} \, = \, g'_{1 \, PP} \, + \, g'_{2 \, PP},
\eeq
where $PP$ can be either $NN$, $\Delta N$ or $\Delta \Delta$.
Obviously, $g'_{1 \, PP}$ ($g'_{2 \, PP}$) is the Landau Migdal
parameter associated with $V_1$ ($V_2$). The $g'_{NN}$
parameter was already
fixed at 0.7, the $g_{NN}$ parameter is fixed at 0.3 and is
completely assign to $V_1$.
Different values for $g'_{\Delta N}$
and $g'_{\Delta \Delta}$ will be considered.

From the numerical calculations it turns out that the appropriate
set of Landau Migdal parameters entering in $V_2$, are the following
$g'_{2 \, NN}$ = 0.5, $g'_{2 \, \Delta N}$ = 0.0 and
$g'_{2 \, \Delta \Delta}$ = 0.3.
This interaction is called $V_a$ in Table I.
As a further simplification all form factors were
evaluated at $q$. In Fig.~8, we compare direct first
and second order in $V_2$ contributions with the full
ring series (where the free structure function was subtracted).
This is done also for $V_b$ interaction.
We see that the agreement is rather good for $V_a$ while it is
poor for $V_b$. In addition, the relative magnitude of the ring
contribution with respect to the free structure function shows that the
disagreement is unacceptable for $V_b$, specially within the
$\Delta$-region.

We have found then an interaction $V_a$, for which its
two first direct perturbative terms account for the full
ring series. Next step is to check that the
exchange terms are smaller than the corresponding direct ones.
Let us consider three different contributions, called
$B_{NN}$, $B_{\Delta N}$
and $B_{\Delta \Delta}$,
depending on the hadronic vertex where the external operator
is attached. More explicitly,
$B_{1 \, NN}$
($B_{1 \, \Delta \Delta}$) is the sum of the first and second
(third and fourth) graphs on line
$R_2$ of Fig.~2. Each subindex 1, indicates that those
are first order contributions. Only $V_{NN}$ contribute to
$B_{1 \, NN}$ and similarly $V_{\Delta N}$ ($V_{\Delta \Delta}$) to
$B_{1 \,\Delta N}$ ($B_{1 \, \Delta \Delta}$).
In higher order terms to
$B_{NN (\Delta N, \Delta \Delta)}$ all interactions
$V_{NN}$, $V_{\Delta N}$ and $V_{\Delta \Delta}$ are present (this can
be easily seen in Eq.~(\ref{eq:r1})).
Before going on,
we must say that the $B_{\Delta N}$ contribution has
very particular features which deserves special attention. For
this reason we discuss now
$B_{NN}$ and $B_{\Delta \Delta}$ and then we go back to
$B_{\Delta N}$.

In Table II, we show direct and exchange first order contributions
$B_{1 \,NN}$ and $B_{1 \, \Delta \Delta}$ to the
RPA series
\footnote{We have preferred to present only first order contributions.
The extremes of the second order ones occurs at different energy
values, which would increase the size of the table with
no additional information.}.
We show results for the two
interactions above mentioned $V_a$ and $V_b$.
First we note that for $V_a$ it holds the condition
that exchange terms keeps smaller than the corresponding
direct ones. This is not the case in $V_b$ for the
$\Delta \Delta$ channel. At this point, we have to
mention that we have used the notation {\it direct} and
{\it exchange} following the standard notation for those graphs.
When the $\Delta$ is involved, direct plus
exchange contributions do not imply antisymmetrization
as the $\Delta$ is a distinguishable particle from nucleons.
At variance with the nucleon sector, exchange terms should
be seen as a particular set of graphs which are present in
the RPA series and eventually can be bigger than the
corresponding direct ones. In fact, there is a compromise
for the values of the parameters entering into the interaction
between the fast convergence of the perturbative terms to the
ring series (shown in Fig.~8) and the condition over exchange terms.

Note that even the starting interaction is the same for both
direct and exchange contributions, for the direct ones
only the central piece of the interaction survives, while
for the exchange contribution both the central and the
tensor term contribute (see for instances, Eqs. (\ref{eq:b1ddd})
and (\ref{eq:b1dd}) in Appendix B).
In particular, this means that the pion contributes
to the exchange terms only.

One of the aims of this work is to study the role
played by exchange terms in the RPA.
All second order exchange contributions
are included and also some higher order ones. We recall
that the residual interaction was split into a contact term
($V_1$), for which exchange terms can be sum up to infinite
order and a finite range piece ($V_2$), for which exchange
terms are considered up to second order. Also cross exchange
terms between $V_1$ and $V_2$ are included up to second order
in $V_2$ and infinite order in $V_1$. In \cite{kn:ba96} we have
discussed an alternative scheme, in which direct contributions
are considered up to infinite order through the RA and
exchange ones are added up to second order. In that work, we saw
that this is not a good approximation for the $NN$ sector. Also,
from table II, it is observed that the situation is even worse
for the $\Delta \Delta$ sector where for $V_b$ exchange contributions
are bigger than the corresponding direct ones.

We turn now to $B_{\Delta N}$ which is dominated by
$V_{\Delta N}$. As mentioned in the last section, no
contact term in the residual interaction which connects
$P_N$ and $P_{\Delta}$ spaces survives. That means that the
direct contact contribution (proportional to $g'_{\Delta N}$) is
exactly cancel by the corresponding exchange one. Then
only the finite range piece of the interaction contribute.
Note that the advantage of our method to evaluate exchange
terms relays on the fact that exchange terms from contact
interaction can be sum up to infinite order. But when
contact terms exactly cancels, there is no way out but
to evaluate each exchange term perturbatively. And in
practice, we can evaluate them up to second order.

Fortunately, the particular behavior of $B_{\Delta N}$
allows a perturbative treatment up to second order. First,
for $V_a$ ($g'_{\Delta N}=0$), the two first direct terms
reproduce the RA almost exactly. Secondly, $B_{1 \, \Delta N}$
has two extremes where each one coincides with the maximum of the
free structure function for the nucleon and delta peaks. In Table III
we show first and second order contributions to $B_{\Delta N}$.
One striking feature is that exchange terms are bigger than
the corresponding direct ones. This is because of the
tensor term of the interaction which contribute to the exchange
term only. In column $exc \, (w.t.)$ of the same table, we present
exchange terms without the tensor contribution. In this case,
exchange terms are smaller than the corresponding direct ones.

In Fig. 9 we present direct plus exchange contribution to $B_{\Delta N}$.
In this figure and Table III, note that while direct
contribution to $B_{\Delta N}$ is positive in the $NN$ sector, the
exchange terms makes the whole contribution to be negative.
Also, in the figure we show the RA result for several values of
$g'_{\Delta N}$. From Appendix A, we see that for $g'_{NN}$,
$g_{NN}$ and $g'_{\Delta \Delta}$ direct plus exchange contact terms
can be account by a re-definition of these parameters. That means
that if one wants to adjust $g'_{NN \, (\Delta \Delta)}$ and
$g_{NN}$ by reproducing any experimental process,
one should take only direct matrix elements for them. But
this is not the case for $g'_{\Delta N}$ where the exchange
term exactly cancel the direct one. Surprisingly, the
RA with a non zero $g'_{\Delta N}$ value reproduce
reasonably well the behavior of the RPA with the
$V_a$ interaction in the $NN$ sector. In the first case
it is $g'_{\Delta N}$ which dominates, while in the
second case it is the tensor force (not present in RA).
It is widely accepted that these kind of correlations with
the $\Delta(1232)$ reduce
the intensity of the transverse quasi elastic structure function. Our
result confirm this, but for a very different reason.

Going back to our scheme, from Table III we observe that
exchange term are considerably reduced from first to second
order in the $NN$ sector. The situation is not so good in the
$\Delta \Delta$ sector. Even thought, $B_{\Delta N}$ is small
in comparison with the free structure function in the $\Delta \Delta$ sector.

Let us resume these considerations about the applicably of the
scheme. We have found an interaction $V_a$ for which its firsts two
direct perturbative terms accounts for the whole RA
and whose exchange terms are smaller than the corresponding
direct ones for $B_{NN}$ and $B_{\Delta \Delta}$.
Even this does not hold for the exchange $B_{\Delta N}$ term,
we will evaluate it up to second order
as this term is small in comparison with the
free structure function.
Going back to Eq.~(\ref{eq:in1}) we see
that $V_2$ (whose value is $V_a$ for the present calculation),
fix the finite range piece of the interaction.
We can vary freely the contact $V_1$ interaction to
construct $V$. As already mentioned,
the only remaining constrain is that $V_1$ should
allows a perturbative treatment.

Now we consider final results for the nuclear structure function. In
Fig.~10, we study the effect of the $\Delta$ over the nuclear
structure function in the $NN$ sector. We show both RPA results with and
without the $\Delta$. The Landau Migdal parameters are the ones
named by $V_b$ ($V_b=V_1+V_a$)
in Table I. Formally, $V_{\Delta \Delta}$ has
an effect on this sector, but the numerical analysis shows
that this is negligible small. It is $V_{\Delta N}$ which
dominates the $\Delta$ contribution over this sector. As
no $g'_{\Delta N}$ parameter appears, we show results for only
one interaction. We see that the $\Delta$ reduce the RPA structure function
in an appreciable way. To get a better understanding of the effect
of RPA correlations over the structure function, we define the quantity,
\beq
\gamma(\bold{q},\hbar \omega)= 100 \;
\frac{R_{PP}^{RPA}(\bold{q},\hbar \omega) \; - \;
R_{PP}^0(\bold{q},\hbar \omega)}
{R_{PP}^0(\bold{q},\hbar \omega)}
\label{eq:gamrp},
\eeq
which is displayed in Fig.~10~b. The $\Delta$ reduce the intensity
of the quasi elastic peak. While RPA correlations without the
$\Delta$ produce a redistribution of the intensity within the $NN$
sector keeping the energy weight sum rule unchanged,
$V_{\Delta N}$ translates intensity from the $NN$- to the
$\Delta \Delta$-sector (as already shown in Fig.~9). Obviously,
the energy weight sum rule is also unchanged but within the
full energy region.

In Fig. 11, we present the RPA structure function for several residual
interaction for both the $NN$ and $\Delta \Delta$-sectors. We
vary only $g'_{\Delta \Delta}$ from 0.4 to 0.6. Also, in
Fig. 11 b, we show $\gamma(\bold{q},\hbar \omega)$.
The effect of RPA correlations is analogous in both sectors. Note
that the changes in $V_{\Delta \Delta}$ makes no appreciable
change in the $NN$-sector. Also, as $g'_{\Delta \Delta}$
grows, this more repulsive interaction moves the
delta peak towards higher energies.
Small changes on the $g'_{\Delta \Delta}$ value makes
small changes in the RPA structure function.

In Fig. 12, we compare the RPA result with the
RA one (that its, is corresponding direct terms). For
RA, we have used $g'_{\Delta N}=0.43$ induced by Fig. 9.
From Fig. 12, we note first that also for direct terms, changes
in $g'_{\Delta \Delta}$ produce no appreciable changes
in the $NN$ sector. But the striking point is the extreme
sensibility of the RA result over the $g'_{\Delta \Delta}$ value
in the $\Delta \Delta$ sector. At variance the RPA structure function
has a smooth behavior. This is a consequence or exchange
terms, which happens to be very important. This is
already suggest by Eq.~(\ref{eq:g1dde}) for instance,
where for a contact interaction the exchange term
reduce the $g'_{\Delta \Delta}$ parameter in about 70 $\%$.

Finally, note that the RA approximation with $g'_{\Delta N}$
and $g'_{\Delta \Delta}$ parameters gives a reasonable
result only in the $NN$ sector, although it is
questionable how to fix $g'_{\Delta N}$.

\subsection{Self Energy results:}

We go on analyzing the structure function per unit volume
at momentum transfer $q=410$ MeV/c by including self energy
insertions. In Fig.~13, we show
the imaginary part of the self energy. Notice
that the quantity we have called self energy is in fact
part of a full set of diagrams and through an average
procedure it depends upon the momentum and energy of the
external probe. It is at variance with other
calculations where self energy insertions depends
on the variables of the particle (hole or $\Delta$),
where it is attached (see refs.
\cite{kn:je93, kn:os87, kn:gi97}).

In Fig.~13 a), we have plotted $Im \, \Sigma^{P_N Q P_N}$
for $V_c$ from Table I.
The non vanishing values for $Im \, \Sigma^{P_N Q_4 P_N}$
lays outside the energy region of interest. From all
self energy insertions $\Sigma^{P_N Q_1 P_N}$ is the most
widely studied. Our result is in agreement with other
calculations where an empirical optical potential is used
(see ref. \cite{kn:co88} for instance). Within a more
complex scheme, analogous correlations are considered in
the optical Green function approach \cite{kn:ho80}.
Self energy for the $\Delta$ are presented in Fig. 13 b). We
have a qualitative agreement with the existing values
for $Im \, \Sigma^{P_{\Delta} Q_1 P_{\Delta}}$ from refs.
\cite{kn:je93} and \cite{kn:os87}, even a rigorous
comparison is not possible due to the
particularities of the average procedure we have mentioned.
Also in Fig 13 a) and b), we present results for
$Im \, \Sigma^{P_{N (\Delta)} Q_{2,3} P_{N (\Delta)}}$, which
are shown for completeness. We concentrate now on
$\Sigma^{P_{N (\Delta)} Q_1 P_{N (\Delta)}}$.

With respect to the exchange terms for both
$\Sigma^{P_N Q_1 P_N}$ and $\Sigma^{P_{\Delta} Q_1 P_{\Delta}}$,
they are shown in Table IV. At variance with RPA correlations,
exchange contributions are small. In
addition, due to the different kind of sums over spin and
isospin, no cancellation occurs for $g'_{\Delta N}$.

In Fig. 14, we study the dependence of
$Im \, \Sigma^{P_{\Delta} Q_1 P_{\Delta}}$ over some
of the parameters entering in our scheme. In Fig. 14 a),
we vary the contact Landau Migdal $g'_{\Delta N}$ parameter,
keeping fixed all other parameters and constants. We
see that $Im \, \Sigma^{P_{\Delta} Q_1 P_{\Delta}}$
is rather sensible to the $g'_{\Delta N}$ value.
In Fig. 14 b), we analyzed different values for the
meson-delta-nucleon cut offs, which are more
uncertain than the corresponding meson-nucleon-nucleon
ones. Without form factors, self energy insertions
would be divergent. Thus the election of the cut off is
crucial in the evaluation of the self energy.
At a first glance, curves b1-b3 may be seeing as
contradictory, as the self energy grows for
decreasing values of the cut off. This behavior
is a consequence of having chosen
$\Lambda_{\pi \Delta N} = \Lambda_{\rho \Delta N}$.
In the analogous expression of Eqs.~(\ref{eq:gpt}) and (\ref{eq:hpt})
for the $\Delta N$ channel we see that there is a competition
of the $\rho$-meson and $g'$ and the $\rho$-meson and
the pion in $\tilde{g'}$ and $\tilde{h'}$, respectively.
Due to the different masses of the pion and the rho
meson, a strong reduction on the rho meson interaction
occurs as the cut off decrease while this reduction
is less pronounce for the pion. That is, even all
terms are reduced, the significant reduction
of the rho meson makes the whole interaction stronger.
Curve b4, where we have taken a different value for
$\Lambda_{\pi \Delta N}$ and $\Lambda_{\rho \Delta N}$,
shows an important reduction.

Before going on analyzing the structure function which
results from diagonal self energy insertions, let us
consider non diagonal contributions. Non diagonal
self energy insertions where evaluate up to second
order in the residual interaction $V_c$. There
are 16 different graphs to it, two of which have
no $\Delta$. Some of them are plotted in Fig.~5.
In Fig. 15 a), we present the contributions
for all non diagonal self energy insertions to the
structure function. These contributions produce a small
redistribution of the intensity. No exchange
term were considered. In ref. \cite{kn:ba97} we have showed
that exchange terms are negligible small
for the $NN$ sector. In Fig. 15 b), we see that this contribution
is not significant.

In Fig. 16, we show the results for the diagonal
self energy insertion over the structure function. The general
effect of self energy insertions is to produce a redistribution
of the intensity from the low to the high energy region for
both the $NN$ and $\Delta \Delta$ sectors. The
peaks are reduced in intensity and shifts to lower energies,
while intensity appears in the dip region and also at
energies beyond the $\Delta \Delta$ peak. In this figure
we show results for both
$\Sigma^{P_{N (\Delta)} Q_1 P_{N (\Delta)}}$ and
$\Sigma^{NN (\Delta \Delta)}$
(see eqs. (\ref{eq:senn}) and (\ref{eq:sedd})).
Note that even $Im \, \Sigma^{P_{\Delta} Q_4 P_{\Delta}} $ is
zero within the energy region of interest,
the same do not holds for its real part. It is not
easy to established which are the appropriate values for the self
energies when the $\Delta$ is involved. As we
have shown in Fig. 14, the self energy depends strongly
on the interaction (and in particular on its cut offs
for the mesons form factors).
It means that within the uncertainty in the election
of the interaction, strongly different results for the self
energy can be obtained.
The aim of this work is to present the scheme and analyzed
the general features of each contribution. For
$\Sigma^{P_{N (\Delta)} Q_{2,3,4} P_{N (\Delta)}}$ self
energy insertions, a deeper analysis for different
interactions is desirable. Due to this and due
to the fact that only $\Sigma^{P_{N (\Delta)} Q_1 P_{N (\Delta)}}$
exchange terms where evaluated, we prefer to take only
the $\Sigma^{P_{N (\Delta)} Q_1 P_{N (\Delta)}}$ self
energy as our final result.
The effect of the $\Delta$ over the $NN$ quasi elastic peak
is not significant at low energies.
For increasing values of the energy it becomes more important
producing an increase of the intensity, being
very relevant in the dip region.

Finally, in Fig. 17 we compare our final result of Eq.~(\ref{eq:rpptot})
with data at momentum transfer $q=410$ and 550 MeV/c.
In this case, we have plotted the transverse response
function for $^{40}Ca$. It is not our intention to reproduce the
data. But we wanted to include this figure in order
to see how do the different contributions stemming from
our scheme compares with the experimental points. We
consider this result encouraging as the interaction chosen
is very simple, in particular the set of cut offs and coupling
constants when the $\Delta$ is involved. And also because other
mechanisms, like ground state correlations beyond RPA ($2p-2h$
states) and MEC add intensity
in this region. Note the behavior around the dip,
only the $\Delta$ can produce the change in the slope
at high energies suggested by data. As mentioned, additional
intensity is provided by $2p-2h$ and MEC.

\section{Conclusions}
In this work we have addressed the role played by the
$\Delta (1232)$ in the nuclear transverse response for
quasi elastic electron scattering. This was done by including
RPA correlations and self energy insertions. All calculations
were done in nuclear matter.

We have presented a formalism which explicitly includes
the $\Delta$ and exchange terms in the RPA series.
At variance with direct terms which can be sum up
to infinite order, exchange terms should be evaluated perturbatively.
In our scheme we have divided the residual interaction in a
weak finite range plus contact term and the remainder contact one.
As for direct terms of any interaction, exchange terms for a contact
interaction can also be sum up to infinite order. These allows
the inclusion of higher order exchange terms.

The effect of exchange terms is important in both the
nucleon and the $\Delta$ peak sectors. In the $\Delta$ sector
the RA gives a narrow peak not suggested by the double differential
cross section for $(e,e')$. The addition of exchange terms
produce a curve which shape is analogous to the one in the
nucleon sector.
Also the role of interference terms between
the nucleon and the $\Delta$ sectors was clarified. Exchange
terms cancel the contact $g'_{\Delta N}$ Landau Migdal parameter.
The result for these interference terms is governed by the
tensor piece of the $V_{\Delta N}$ interaction, given a redistribution
of the intensity from the nuclear to the $\Delta$ sector.

The final result for RPA correlations is to redistribute
the intensity from the low to the high energy region, keeping
the energy weight sum rule unchanged. The direction towards the
intensity moves is a consequence of the repulsive character of
the interaction.

We have considered self energy insertions which dressed a single
$ph$ and $\Delta h$ bubble. They were divided into two sets,
depending on which bubble they act and then, subdivided into four
terms depending on the intermediate states
(see eqs. (\ref{eq:senn}) and (\ref{eq:sedd})).
Sums over spin and isospin are very different from
the RPA ones and so the integration over the internal momentum.
For direct self energy insertions the tensor
term of the interaction is present, no cancellation of
$g'_{\Delta N}$ occurs and exchange terms are unimportant.

Self energy insertions produce also a redistribution of the
intensity from the low to the high energy regions. But
while RPA correlations makes this redistribution within
$ph$ and $\Delta h$ states, the self energy opens new
channels like $2p-2h$ and higher order states. This
is specially important for the dip region, even a complete
description should include ground state correlations beyond
RPA and MEC. Finally, self energy insertions are very
sensitive to the residual interaction employed.

As a final remark, let us mentioned that
we have preferred not to compare our result with the double
differential cross section for $(e,e')$. The reason for this is twofold.
First, the longitudinal response function should be included
and discussed. In this sense, the residual interaction should
be modified. Secondly, this double differential cross
section could be a tool to study deeper the
residual interaction for the $\Delta$,
which is beyond the scope of the present work.

\section*{Acknowledgments}

I would like to thank O. Civitarese for fruitful
discussions and for the critical reading of the
manuscript.
Also I would like to thank C. F. Williamson for
communicating the $^{40}$Ca experimental points of the
MIT-Bard College-Louisiana State University-Northwestern
University-Ohio University collaboration.
This work has been partially supported
by the Agencia Nacional de Promoci\'on Cient\'{\i}fica
y Tecnol\'ogica,
under contract PMT-PICT-0079.

\newpage

\section*{APPENDIX A}
In this appendix we show how to modified Landau Migdal parameters
to account for exchange term. The direct $V_1$ interaction for
each $P$-channel needed in the RPA, are,
\beqar
V_{1 \; NN} & = & \frac{f_{\pi NN}^2}
{\mu_{\pi}^2}
(g_{1 \; NN} \: \mbox{\boldmath $\sigma \cdot \sigma ' $} +
g'_{1 \; NN} \mbox{\boldmath $\tau \cdot \tau ' $} \,
\mbox{\boldmath $\sigma \cdot \sigma ' $}  )
\label{eq:v1nn}, \\ \nonumber \\
V_{1 \; \Delta N} & = & \frac{f_{\pi NN} f_{\pi \Delta N}}
{\mu_{\pi}^2}  \; \;
g'_{1 \; \Delta N}  \;  \; \mbox{\boldmath $\tau \cdot T' $} \,
\mbox{\boldmath $\sigma \cdot S' $}
\label{eq:v1nd}, \\ \nonumber \\
V_{1 \; \Delta \Delta} & = & \frac{f_{\pi \Delta N}^2}
{\mu_{\pi}^2}  \; \;
g'_{1 \; \Delta \Delta} \; \; \mbox{\boldmath $\tau \cdot {\cal T}' $}
\, \mbox{\boldmath $\sigma \cdot {\cal S}' $}
\label{eq:v1dd}.
\eeqar
There is two ways to obtained the exchange terms. One is to act with
exchange operators $P_{\sigma}$ and $P_{\tau}$ over the direct
interaction (see for instance the appendix of ref. \cite{kn:sh89}),
then,
\beq
(V_{1 \; NN})_E = - P_{\sigma}	P_{\tau} V_{1 \; NN}
\label{eq:v1ex},
\eeq
given the following values for the Landau Migdal parameters,
\beqar
(g_{1 \; NN})_E & = & \frac{1}{4} \; (g_{1 \; NN} \; + \; 3 g'_{1 \; NN})
\label{eq:g1nn}, \\ \nonumber \\
(g'_{1 \; NN})_E & = & \frac{1}{4} \; (g_{1 \; NN} \; - \; g'_{1 \; NN})
\label{eq:gp1nn}.
\eeqar
Finally, the values needed to evaluate $R_1$ are,
$(g_{1 \; NN})_{D+E}=g_{1 \; NN}+(g_{1 \; NN})_E$ and
$(g'_{1 \; NN})_{D+E}=g'_{1 \; NN}+(g'_{1 \; NN})_E$.

We describe now an
alternative and fully equivalent method \cite{kn:ho86}.
Let us consider the $V_{1 \; \Delta \Delta}$.
We evaluate graph $D$
and $E$ of Fig. 18 using $V_{1 \; \Delta \Delta}$.
To evaluate $(g'_{1 \; \Delta \Delta})_{D+E}$ one has to
put matrices $\mbox{\boldmath $T$}$ and $\mbox{\boldmath $S$}$
in the external vertices to make sums over spin and
isospin, respectively.
Note that the mesonic vertices
are different between $D$ and $E$. After performing sums
over spin and isospin we get,
\beq
(g'_{1 \; \Delta \Delta})_{D+E} = \{ g'_{1 \; \Delta \Delta} \; - \;
\frac{225}{64} \; \frac{f_{\pi NN} f_{\pi \Delta \Delta}}
{f_{\pi \Delta N}^2} \;
g'_{1 \; \Delta \Delta} \}
\label{eq:g1dd}.
\eeq
If we use, $f_{\pi \Delta N} = 2 f_{\pi NN}$
and  $f_{\pi \Delta \Delta} = \frac{4}{5} f_{\pi NN}$ we obtain,
\beq
(g'_{1 \; \Delta \Delta})_{D+E} = \frac{19}{64} \;
g'_{1 \; \Delta \Delta}
\label{eq:g1dde}.
\eeq
In an analogous way, it can be seen that
$(g'_{1 \; \Delta N})_{D+E} = 0$; a result already known
\cite{kn:to87}.

\newpage

\section*{APPENDIX B}

In this Appendix, we present explicit expressions for the
different graphs needed to build up our antisymmetric
RPA for nuclear matter. We do not reproduce here the $ph$  contributions
as they were already published \cite{kn:ba96}.

First, we need to define the $\Delta$ propagator,
\beq
G_{\Delta}(p) = \frac{1}{p_0 - \frac{\bold{p}}{2 m_{\Delta}}
-\delta m + i \eta}
\eeq
where $m_{\Delta}$ is the $\Delta(1232)$ mass and $\delta m$
is the mass difference between $\Delta$ and nucleon.

It is convenient to define now two function $x$ and $z$ related with
the $ph$ and $\Delta h$ bubble, as,
\beq
x(\bold{Q}, \nu)= \int d^3 h \frac{ \theta (| \bold{h} + \bold{Q} | - 1)
\theta ( 1 - h) }{\nu - (Q^2/2 + \bold{h.Q}) + i \eta} \; - \;
\int d^3 h \frac{ \theta (| \bold{h} + \bold{Q} | - 1)
\theta ( 1 - h) }{\nu + (Q^2/2 - \bold{h.Q})}
\eeq
and
\beq
z(\bold{Q}, \nu)= \int d^3 h \frac{\theta ( 1 - h) }
{\nu - (\frac{c-1}{2} h^2 + c \; \bold{h.Q}) + \delta + i \eta} \; - \;
\int d^3 h \frac{\theta ( 1 - h) }
{\nu + (\frac{c-1}{2} h^2 - c \; \bold{h.Q}) + \delta }
\eeq
where $c=m/m_{\Delta}$ and $\delta= \delta m/(2 \varepsilon_F)$
We have used dimensionless quantities as described in the text.

We show now first order contributions in $V_2$.
The external probe can create (or destroyed) a $ph$ or
$\Delta h$ pair. The interference terms between these two pairs are,
\beqar
(B_{1 \; \Delta N} (\bold{Q}, \nu))_D  & = & \frac{8}{3}
\frac{1}{(2 \pi)^3}\
\frac{A}{\hbar c \mu_{\pi}^2 \; k_F}
(\frac{f_{\pi NN} f_{\pi \Delta N}}{4 \pi \hbar c } \ ) \;
Q^2 \; \mu_v \; \mu_{\Delta} \; G_E \; G_{\Delta}
\nonumber \\ [3.mm]
& & \Gamma_{\pi NN}(Q) \; \Gamma_{\pi \Delta N}(Q) \;
\tilde{g}'_{2 \Delta N}(Q) \; x(\bold{Q}, \nu) \; z(\bold{Q}, \nu)
\label{eq:b1ndd}
\eeqar
and
\beqar
(B_{1 \; \Delta N} (\bold{Q}, \nu))_E & = & - \frac{1}{6}
\frac{1}{(2 \pi)^3}\
\frac{A}{\hbar c \mu_{\pi}^2 \; k_F}
(\frac{f_{\pi NN} f_{\pi \Delta N}}{4 \pi \hbar c } \ ) \;
Q^2 \; \mu_{\Delta} \; \mu_v \; G_E \; G_{\Delta}
\nonumber \\ [3.mm]
& & \int d^3 h \ \int d^3 k \ \theta ( 1 - h)
\theta (1 - | \bold{h} + \bold{k} | )
\theta (| \bold{h} + \bold{Q} + \bold{k} | - 1)
\nonumber \\ [3.mm]
& & \;	\Gamma_{\pi NN}(k) \; \Gamma_{\pi \Delta N}(k) \;
\; \{ 8 \tilde{g}'_{2 \Delta N} + \tilde{h}'_{\Delta N} \;
(3 - (\bold{\widehat{k} . \widehat{Q}})^2) \}
\nonumber \\ [3.mm]
& &
\{ \frac{1}{\nu - \alpha_1 + i \eta} \, - \,
\frac{1}{\nu + \alpha_1 } \} \;
\{ \frac{1}{\nu - \alpha_2 + i \eta} \, - \,
\frac{1}{\nu + \alpha_2 } \}
\nonumber \\ [3.mm]
\label{eq:b1nd}
\eeqar
Expressions for the graphs shown in the third and fourth
places of line $R_2$ of Fig.~2, are,
\beqar
(B_{1 \; \Delta \Delta} (\bold{Q}, \nu))_D  & = & \frac{32}{27}
\frac{1}{(2 \pi)^3}\
\frac{A}{\hbar c \mu_{\pi}^2 \; k_F}
(\frac{f_{\pi NN} f_{\pi \Delta N}}{4 \pi \hbar c } \ ) \;
Q^2 \; \mu_{\Delta}^2 \; G_E \; G_{\Delta}
\nonumber \\ [3.mm]
& & \Gamma_{\pi NN}(Q) \; \Gamma_{\pi \Delta N}(Q) \;
\tilde{g}'_{2 \Delta N}(Q) \; (z(\bold{Q}, \nu))^2
\label{eq:b1ddd}
\eeqar
and
\beqar
(B_{1 \; \Delta \Delta} (\bold{Q}, \nu))_E & = & - \frac{5}{3}
\frac{1}{(2 \pi)^3}\
\frac{A}{\hbar c \mu_{\pi}^2 \; k_F}
(\frac{f_{\pi NN} f_{\pi \Delta \Delta}}{4 \pi \hbar c } \ ) \;
Q^2 \; \mu_{\Delta}^2 \;  (G_{\Delta})^2
\int d^3 h \ \int d^3 k \
\nonumber \\ [3.mm]
& & \theta ( 1 - h)
\theta (1 - | \bold{h} + \bold{k} | )
\; \Gamma_{\pi \Delta N}^2 (k) \;
\; \{ 10 \tilde{g}'_{2 \Delta \Delta} + \tilde{h}'_{\Delta \Delta} \;
(3 + (\bold{\widehat{k} . \widehat{Q}})^2) \}
\nonumber \\ [3.mm]\
& &
\{ \frac{1}{\nu - \alpha_3 + i \eta} \, - \,
\frac{1}{\nu + \alpha_3 } \} \;
\{ \frac{1}{\nu - \alpha_4 + i \eta} \, - \,
\frac{1}{\nu + \alpha_4 } \}
\nonumber \\ [3.mm]
\label{eq:b1dd}
\eeqar
where we have defined,
\beq
\tilde{g}'_{2 \Delta N} (k) =
g'_{2 \Delta N}-
\frac{\Gamma_{\rho \Delta N}^2 (k)}
{\Gamma_{\pi \Delta N}^2 (k)}\
C_{\rho \Delta N}
\frac{k^2}{k^2 + \mu_{\rho}^2} \
\eeq
with an analogous expression for $\tilde{g}'_{2 \Delta \Delta}$ and
\beqar
\alpha_1 & = & \frac{c-1}{2} h^2 + \frac{c}{2} Q^2
+ c \; \bold{h.Q} + \delta,
\nonumber \\ [3.mm]\
\alpha_2 &= & Q^2/2 + \bold{k.Q} + \bold{h.Q}
\nonumber \\ [3.mm]\
\alpha_3 & = & \alpha_1
\nonumber \\ [3.mm]\
\alpha_4 & = & \frac{c-1}{2} h^2 + \frac{c-1}{2} k^2
+ \frac{c}{2} Q^2 + (c-1) \; \bold{h.k} +
c \; \bold{h.Q} + c \; \bold{k.Q}+ \delta.
\eeqar
Finally, $B_{1 \; \Delta N}=(B_{1 \; \Delta N})_D+
(B_{1 \; \Delta N})_E $ and
$B_{1 \; \Delta \Delta}=(B_{1 \; \Delta \Delta})_D+
(B_{1 \; \Delta \Delta})_E $.

For second order contributions, we show exchange contributions
only. It is \\
straightforward to obtain direct ones. We need
to present the expression for graphs $B_{1 NN}$,
\beqar
(B_{1 \; NN} (\bold{Q}, \nu))_E & = & - \frac{1}{8}
\frac{1}{(2 \pi)^3}\
\frac{A}{\hbar c \mu_{\pi}^2 \; k_F}
(\frac{f_{\pi NN}^2}{4 \pi \hbar c } \ ) \;
Q^2 \; ({\mu_v}^2 - 3 {\mu_s}^2 ) \;  (G_E)^2
\int d^3 h \ \int d^3 k \
\nonumber \\ [3.mm]
& & \theta ( 1 - h)
\theta (| \bold{p} + \bold{Q} | - 1)
\theta (1 - | \bold{h} + \bold{k} | )
\theta (| \bold{h} + \bold{Q} + \bold{k} | - 1)
\nonumber \\ [3.mm]\
& &
\; \Gamma_{\pi NN}^2 (k) \;
\; \{ \tilde{g}'_{2 NN} + \tilde{h}'_{NN} \;
((\bold{\widehat{k} . \widehat{Q}})^2 \}
\nonumber \\ [3.mm]\
& &
\{ \frac{1}{\nu - \alpha + i \eta} \, - \,
\frac{1}{\nu + \alpha } \} \;
\{ \frac{1}{\nu - \alpha' + i \eta} \, - \,
\frac{1}{\nu + \alpha' } \}
\nonumber \\ [3.mm]
\label{eq:b1nn}
\eeqar
with
\beqar
\alpha & = & Q^2/2 + \bold{h.Q}
\nonumber \\ [3.mm]
\alpha' & = & Q^2/2 + \bold{k.Q} + \bold{h.Q}
\eeqar
Also, we define the functions,
\beqar
\eta_{NN}(Q) & = &
\frac{m}{(2 \pi)^2}\
(\frac{f_{\pi NN}^2}{4 \pi \hbar c } \ ) \;
\frac{\tilde{g}'_{2 NN} (Q)}{\hbar c \mu_{\pi}^2 \; k_F}
\nonumber \\ [3.mm]
\eta_{\Delta N}(Q) & = &
\frac{m}{(2 \pi)^2}\
(\frac{f_{\pi NN} f_{\pi \Delta N}}{4 \pi \hbar c } \ ) \;
\frac{\tilde{g}'_{2 \Delta N} (Q)}{\hbar c \mu_{\pi}^2 \; k_F}
\nonumber \\ [3.mm]
\eta_{\Delta \Delta }(Q) & = &
\frac{m}{(2 \pi)^2}\
(\frac{f_{\pi \Delta N}^2}{4 \pi \hbar c } \ ) \;
\frac{\tilde{g}'_{2 \Delta \Delta } (Q)}{\hbar c \mu_{\pi}^2 \; k_F}
\eeqar
Now second order exchange contributions are given by,
\beqar
B_{2 NNN} & = & 8 \; \eta_{NN} \; B_{1 \; NN} \; x(\bold{Q}, \nu),
\nonumber \\ [3.mm]
B_{2 NN \Delta} & = & \frac{64}{9} \; \eta_{N \Delta} \;
B_{1 \; NN} \; z(\bold{Q}, \nu),
\nonumber \\ [3.mm]
B_{2 N \Delta N} & = & 8 \; \eta_{N \Delta} \;
B_{1 \; \Delta N} \; x(\bold{Q}, \nu),
\nonumber \\ [3.mm]
B_{2 \Delta NN} & = & 8 \; \eta_{N \Delta} \;
B_{1 \; \Delta N} \; x(\bold{Q}, \nu),
\nonumber \\ [3.mm]
B_{2 N \Delta \Delta} & = & \frac{64}{9} \; \eta_{\Delta \Delta} \;
B_{1 \; \Delta N} \; z(\bold{Q}, \nu),
\nonumber \\ [3.mm]
B_{2 \Delta N \Delta} & = & \frac{64}{9} \; \eta_{\Delta N} \;
B_{1 \; \Delta N} \; z(\bold{Q}, \nu),
\nonumber \\ [3.mm]
B_{2 \Delta \Delta N} & = & 8 \; \eta_{\Delta N} \;
B_{1 \; \Delta \Delta} \; x(\bold{Q}, \nu),
\nonumber \\ [3.mm]
B_{2 \Delta \Delta \Delta} & = & \frac{64}{9} \; \eta_{\Delta \Delta} \;
B_{1 \; \Delta \Delta} \; z(\bold{Q}, \nu),
\eeqar

\newpage

\section*{APPENDIX C}

In this appendix we show explicit expressions for
the self-energy insertions
Even direct contributions in the $ph$ sector
can be found in \cite{kn:ba95} and exchange ones are presented
in \cite{kn:ba97}, we reproduce them for completeness.

In Eqs. (\ref{eq:senn}) we have,
\beqar
\Sigma^{P_N Q_1 P_N} (\bold{h} , \bold{Q} , \nu) & = &
\frac{3}{2 \pi^4}
(\frac{f_{\pi NN}^2}{4 \pi \hbar c} \ )^2
\frac{m c^2 \; {k_F}^4}{\mu_{\pi}^4} \ \int d^3 l \ \int d^3 p \
\Gamma_{\pi NN}^4 (l)
\theta (| \bold{p} + \bold{l}/2 | - 1)	\nonumber \\ [3. mm]
& &
\theta (1 - | \bold{p} - \bold{l}/2 | ) \;
( g_{NN}^2 + 3 (\tilde{g}'_{NN})^2 + (\tilde{h}'_{NN})^2
+ 2 \tilde{g}'_{NN} \tilde{h}'_{NN} ) \nonumber \\ [3. mm]
& & \{ \theta (| \bold{h} + \bold{Q} - \bold{l} | - 1) \;
\frac{1}{\nu - \alpha_1 + i \eta } \; + \; \nonumber \\ [3. mm]
& & \; + \; \theta (1 - | \bold{h} + \bold{l} | ) \;
\frac{1}{\nu - \alpha'_1 + i \eta }  \},
\label{eq:snnq1} \\ [5. mm]
\Sigma^{P_N Q_2 P_N} (\bold{h} ,\bold{q}, \nu) & = &
\frac{2}{3} \,
\frac{m c^2 \, {k_F}^4}{\mu_{\pi}^4 \pi^4}
(\frac{f_{\pi NN} f_{\pi \Delta N}}{4 \pi \hbar c} \ )^2
\int d^3 k \ \int d^3 h' \
\theta (1-h')
\nonumber \\ [3. mm]
& &
\Gamma_{\pi NN}^2 (k) \Gamma_{\pi \Delta N}^2 (k) \;
(3 (\tilde{g}'_{\Delta N})^2 \, + \,
(\tilde{h}'_{\Delta N})^2 \, + \,
2 \tilde{g}'_{\Delta N} \tilde{h}'_{\Delta N} )
\nonumber \\ [3. mm]
& &
\{ \theta (| \bold{h} + \bold{Q} - \bold{k} | - 1) \;
\frac{1}
{\nu \, - \, \alpha_2 \, +  \, i \eta } \; + \;
\nonumber \\ [3. mm]
& &
\theta (1 - | \bold{h} + \bold{k} |)
\frac{1}
{\nu \, - \, \alpha'_2 \, +  \, i \eta } \},
\label{eq:sppq2} \\ [5. mm]
\Sigma^{P_N Q_3 P_N} (\bold{h} ,\bold{q}, \nu) & = &
\frac{2}{3} \,
\frac{m c^2 \, {k_F}^4}{\mu_{\pi}^4 \pi^4}
(\frac{f_{\pi NN} f_{\pi \Delta N}}{4 \pi \hbar c} \ )^2
\int d^3 k \ \int d^3 h' \
\theta (1-h')
\nonumber \\ [3. mm]
& &
\theta (| \bold{h'} + \bold{k} | - 1)
\Gamma_{\pi NN}^2 (k) \Gamma_{\pi \Delta N}^2 (k)
\nonumber \\ [3. mm]
& &
(3 (\tilde{g}'_{\Delta N})^2 \, + \,
(\tilde{h}'_{\Delta N})^2 \, + \,
2 \tilde{g}'_{\Delta N} \tilde{h}'_{\Delta N} ) \;
\frac{1}
{\nu \, - \, \alpha_3 \, +  \, i \eta },
\label{eq:sppq3} \\ [5. mm]
\Sigma^{P_N Q_4 P_N} (\bold{h} ,\bold{q}, \nu) & = &
\frac{8}{27} \,
\frac{m c^2 \, {k_F}^4}{\mu_{\pi}^4 \pi^4}
(\frac{f_{\pi \Delta \Delta }^2}{4 \pi \hbar c} )^2
\int d^3 k \ \int d^3 h' \
\theta (1-h')
\nonumber \\ [3. mm]
& &
\Gamma_{\pi \Delta N}^4 (k)
(3 (\tilde{g}'_{\Delta \Delta })^2 \, + \,
(\tilde{h}'_{\Delta N})^2 \, + \,
2 \tilde{g}'_{\Delta \Delta } \tilde{h}'_{\Delta \Delta } ) \;
\frac{1}
{\nu \, - \, \alpha_4 \, +  \, i \eta } \label{eq:sppq4} \\
\nonumber
\eeqar
and Eqs. (\ref{eq:sedd}),
\beqar
\Sigma^{P_{\Delta} Q_1 P_{\Delta}} (\bold{h} , \bold{Q} , \nu) & = &
\frac{2}{3 \pi^4}
(\frac{f_{\pi NN} f_{\pi \Delta N}}{4 \pi \hbar c} \ )^2
\frac{m c^2 \; {k_F}^4}{\mu_{\pi}^4} \ \int d^3 l \ \int d^3 p \
\Gamma_{\pi NN}^2 (l) \Gamma_{\pi \Delta N}^2 (l)
\nonumber \\ [3. mm]
& &
\theta (| \bold{p} + \bold{l}/2 | - 1) \;
\theta (1 - | \bold{p} - \bold{l}/2 | ) \;
\theta (| \bold{h} + \bold{Q} - \bold{l} | - 1) \nonumber \\ [3. mm]
& &
(3 (\tilde{g}'_{N \Delta})^2
+ (\tilde{h}'_{N \Delta})^2
+ 2 \tilde{g}'_{N \Delta} \tilde{h}'_{N \Delta} )
\frac{1}{\nu - \alpha_1 + i \eta }, \label{eq:sddq1} \\ [5. mm]
\Sigma^{P_{\Delta} Q_2 P_{\Delta}} (\bold{h} ,\bold{q}, \nu) & = &
\frac{8}{27} \,
\frac{m c^2 \, {k_F}^4}{\mu_{\pi}^4 \pi^4}
(\frac{f_{\pi \Delta N}^2}{4 \pi \hbar c} \ )^2
\int d^3 k \ \int d^3 h' \
\theta (1-h')
\nonumber \\ [3. mm]
& &
\Gamma_{\pi \Delta N}^4 (k) \;
(3 (\tilde{g}'_{\Delta \Delta})^2 \, + \,
(\tilde{h}'_{\Delta \Delta})^2 \, + \,
2 \tilde{g}'_{\Delta \Delta} \tilde{h}'_{\Delta \Delta} )
\nonumber \\ [3. mm]
& &
\theta (| \bold{h} + \bold{Q} - \bold{k} | - 1) \;
\frac{1}
{\nu \, - \, \alpha_2 \, +  \, i \eta }, \label{eq:sddq2} \\ [5. mm]
\Sigma^{P_{\Delta} Q_3 P_{\Delta}} (\bold{h} ,\bold{q}, \nu) & = &
\frac{2}{3} \,
\frac{m c^2 \, {k_F}^4}{\mu_{\pi}^4 \pi^4}
(\frac{f_{\pi NN} f_{\pi \Delta \Delta}}{4 \pi \hbar c} \ )^2
\int d^3 k \ \int d^3 h' \
\theta (1-h')
\nonumber \\ [3. mm]
& &
\theta (| \bold{h'} + \bold{k} | - 1)
\Gamma_{\pi NN}^2 (k) \Gamma_{\pi \Delta \Delta}^2 (k)
\nonumber \\ [3. mm]
& &
(3 (\tilde{g}'_{\Delta N})^2 \, + \,
(\tilde{h}'_{\Delta N})^2 \, + \,
2 \tilde{g}'_{\Delta N} \tilde{h}'_{\Delta N} ) \;
\frac{1}
{\nu \, - \, \alpha_3 \, +  \, i \eta }, \label{eq:sddq3} \\ [5. mm]
\Sigma^{P_{\Delta} Q_4 P_{\Delta}} (\bold{h} ,\bold{q}, \nu) & = &
\frac{8}{27} \,
\frac{m c^2 \, {k_F}^4}{\mu_{\pi}^4 \pi^4}
(\frac{f_{\pi \Delta \Delta } f_{\pi \Delta N}}{4 \pi \hbar c} )^2
\int d^3 k \ \int d^3 h' \
\theta (1-h')
\Gamma_{\pi \Delta \Delta}^2 (k)
\nonumber \\ [3. mm]
& &
\Gamma_{\pi \Delta N}^2 (k)
(3 (\tilde{g}'_{\Delta \Delta })^2 \, + \,
(\tilde{h}'_{\Delta N})^2 \, + \,
2 \tilde{g}'_{\Delta \Delta } \tilde{h}'_{\Delta \Delta } ) \;
\frac{1}
{\nu \, - \, \alpha_4 \, +  \, i \eta } \label{eq:sddq4} \\ [5. mm]
\nonumber
\eeqar
where,
\beqar
\alpha_1 & = & Q^2/2 + \bold{h.Q} - \bold{l.
(h + Q - l)}/2 + \bold{p.l}
\nonumber \\ [3.mm]
\alpha'_1 & = & Q^2/2 - l^2/2 + \bold{h.(l - Q)} + \bold{p.l}
\nonumber \\ [3.mm]
\alpha_2 & = & Q^2/2 + \frac{c+1}{2} k^2 + \frac{c-1}{2} h'^2 +
\bold{h.Q} - \bold{k.Q}  -
\bold{h.k} + c \, \bold{k.h'} + \delta
\nonumber \\ [3.mm]
\alpha'_2 & = & Q^2/2 + \frac{c+1}{2} k^2 + \frac{c-1}{2} h'^2 +
\bold{h.Q} - \bold{k.h}
+ c \, \bold{k.h'} + \delta
\nonumber \\ [3. mm]
\alpha_3 & = & \frac{c}{2} Q^2 +
\frac{c+1}{2} k^2 + \frac{c-1}{2} h^2 +
c \, \bold{h.Q} - c \,\bold{k.Q}  -
c \, \bold{h.k} + \bold{k.h'} + \delta
\nonumber \\ [3.mm]
\alpha_4 & = & \frac{c}{2} Q^2 +
c \, k^2 + \frac{c-1}{2} h^2 + \frac{c-1}{2} h'^2 +
c \, \bold{h.Q} + c \,\bold{k.h'}  -
c \, \bold{h.k} - c\, \bold{k.Q} + 2 \, \delta
\nonumber
\eeqar

Exchange terms are given by,
\beqar
\Sigma^{P_N Q_{E1} P_N} (\bold{h} ,\bold{q}, \nu) & = &
- \frac{1}{(2 \pi)^4}\
(\frac{f_{\pi NN}^2}{4 \pi \hbar c} \ )^2
\frac{m c^2 \, {k_F}^4}{\mu_{\pi}^4} \ \int d^3 k \ \int d^3 k' \
\nonumber \\ [3.mm]
& &
\Gamma_{\pi NN}^2 (k) \Gamma_{\pi NN}^2 (k')
\theta (1 - | \bold{h} + \bold{Q} - \bold{k} - \bold{k'} | )
\nonumber \\ [3.mm]
& &
\theta (| \bold{h} + \bold{Q} - \bold{k} | - 1)
\theta (| \bold{h} + \bold{Q} - \bold{k'} | - 1)
\nonumber \\ [3.mm]
& &
(3 \tilde{g}'^2 - ( 2 (\bold{\widehat{k}.\widehat{k'}} )^2 - 1)
 \tilde{h}'^2  + 2 \tilde{g}' \tilde{h}' ) \;
\frac{1}
{\nu - \alpha_{E1} + i \eta }  \
\label{eq:sigphex} \\ [5.mm]
\Sigma^{P_{\Delta} Q_{E1} P_{\Delta}} (\bold{h} ,\bold{q}, \nu) & = &
- \frac{1}{6} \,
\frac{m c^2 \, {k_F}^4}{\mu_{\pi}^4 \pi^4}
(\frac{f_{\pi NN} f_{\pi \Delta N}}{4 \pi \hbar c} \ )^2
\int d^3 k \ \int d^3 k' \
\theta (| \bold{Q} + \bold{h} - \bold{k} | - 1 )
\nonumber \\ [3. mm]
& &
\theta (| \bold{Q} + \bold{h} - \bold{k'} | - 1 )
\theta ( 1 - | \bold{Q} + \bold{h} - \bold{k'} - \bold{k'} |)
\nonumber \\ [3. mm]
& &
\Gamma_{\pi NN} (k) \Gamma_{\pi NN} (k')
\Gamma_{\pi \Delta N} (k) \Gamma_{\pi \Delta N} (k')
\nonumber \\ [3. mm]
& &
\{ 3 \tilde{g}'_{\Delta N}(k) \tilde{g}'_{\Delta N}(k') \, + \,
\frac{3}{2} \ \tilde{g}'_{\Delta N}(k) \tilde{h}'_{\Delta N}(k')
(5 - 3 (\bold{\widehat{k'}.\widehat{Q}} )^2)
\, + \,
\nonumber \\ [3. mm]
& &
\frac{3}{2} \ \tilde{g}'_{\Delta N}(k') \tilde{h}'_{\Delta N}(k)
(5 - 3 (\bold{\widehat{k}.\widehat{Q}} )^2)
\, + \,
\nonumber \\ [3. mm]
& &
\frac{1}{8} \
\tilde{h}'_{\Delta N}(k) \tilde{h}'_{\Delta N}(k') \;
(2+3 (\bold{\widehat{k}.\widehat{h}} )^2  +
2 \bold{\widehat{Q}.(\widehat{k} \times \widehat{k'}} )^2
\nonumber \\ [3. mm]
& &
- (\bold{\widehat{Q}.\widehat{k}} )^2
- (\bold{\widehat{Q}.\widehat{k'}} )^2
- (\bold{\widehat{Q}.\widehat{k}}) (\bold{\widehat{Q}.\widehat{k'}})
  (\bold{\widehat{k}.\widehat{k'}}) \}
\frac{1}
{\nu \, - \, \alpha_{E2} \, +  \, i \eta }  \nonumber \\ \\ [5. mm]
\label{eq:sigddex}
\eeqar
where,
\beqar
\alpha_{E1} & = &  Q^2/2 +
\frac{c-1}{2} h^2 + \frac{c-1}{2} k^2 +  \frac{c-1}{2} k'^2 -
\nonumber \\
& &
(c-1) \, \bold{h.k} + (c-1) \, \bold{h.k'}  +
c \, \bold{k.k'} + \bold{h.q} + \delta
\nonumber
\eeqar

In order to simplify the calculation, it is a good approximation
to eliminate the dependence on the hole momentum, by an average
procedure (see ref. \cite{kn:ba95}), as follows,

\beq
\Sigma^{PQP} (\bold{Q}, \nu) \equiv
\frac{1}{ \frac{4}{3} \ \pi} \ \int d^3 h \
\Sigma^{PQP} (\bold{Q}, \nu, \bold{h})
\eeq

\newpage

\vfill
\eject
\section*{Figure Captions}
\begin{description}

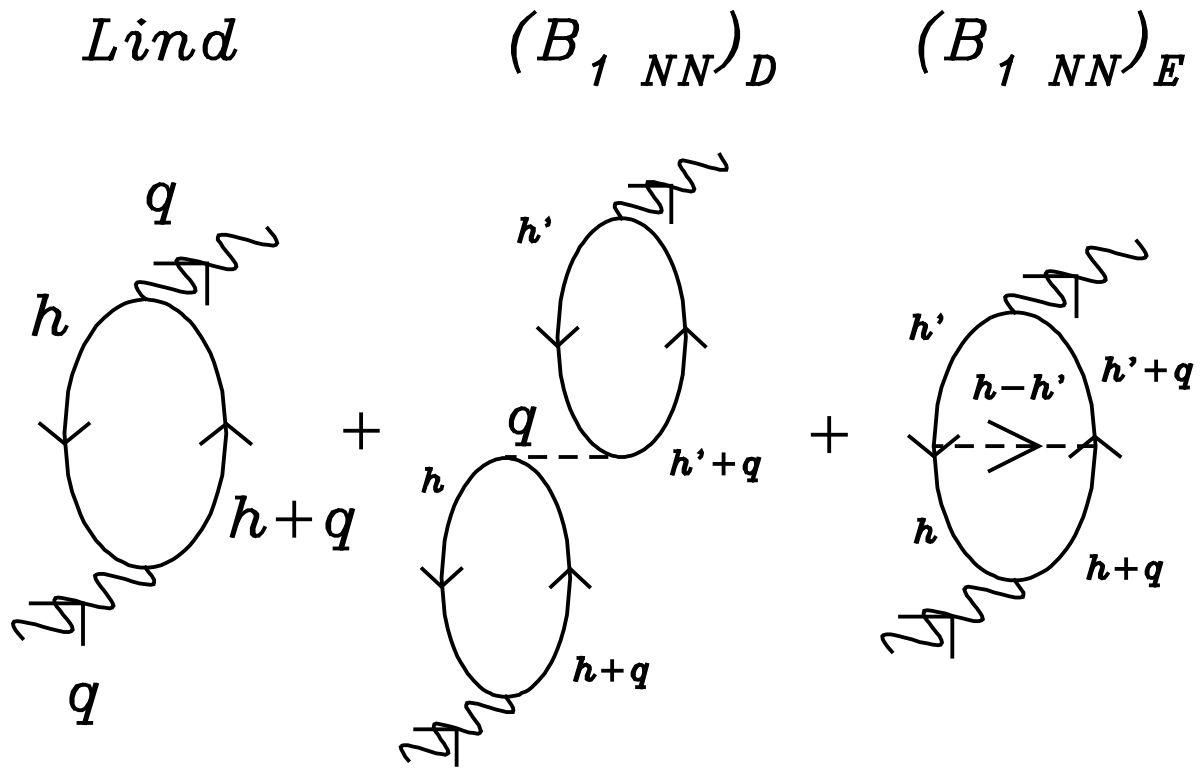
\item [Figure 1:]
Goldstone diagrams stemming from the free response ($Lind$) and
the first order direct (exchange) $ph$ term from the RPA series,
designed by
$(B_{1 \, NN})_D$ ($(B_{1 \, NN})_E$).
In every diagram two wavy lines represent the external
probe with momentum and energy $\bold{q}$ and $\hbar \omega$,
respectively. The dashed line represents the residual interaction.
Only forward-going contributions are shown. We have assigned
values to the momentum carried by each line.

\item [Figure 2:]
Goldstone diagrams from contributions $R_1$, $R_{12}$ and $R_2$
(see Eqs.~(\ref{eq:gw5a})-(\ref{eq:gw5c}).
A dot between bubbles represents the sum of direct plus
exchange contact interaction. Only some first order in
$V_1$ and $V_2$ are shown. A double line represents the delta.

\item [Figure 3:]
Goldstone diagrams stemming from Eq.~(\ref{eq:r12}).
Bubbles with a straight line crossing in the middle represent
the sum of direct and exchange first order in $V_2$ contributions.

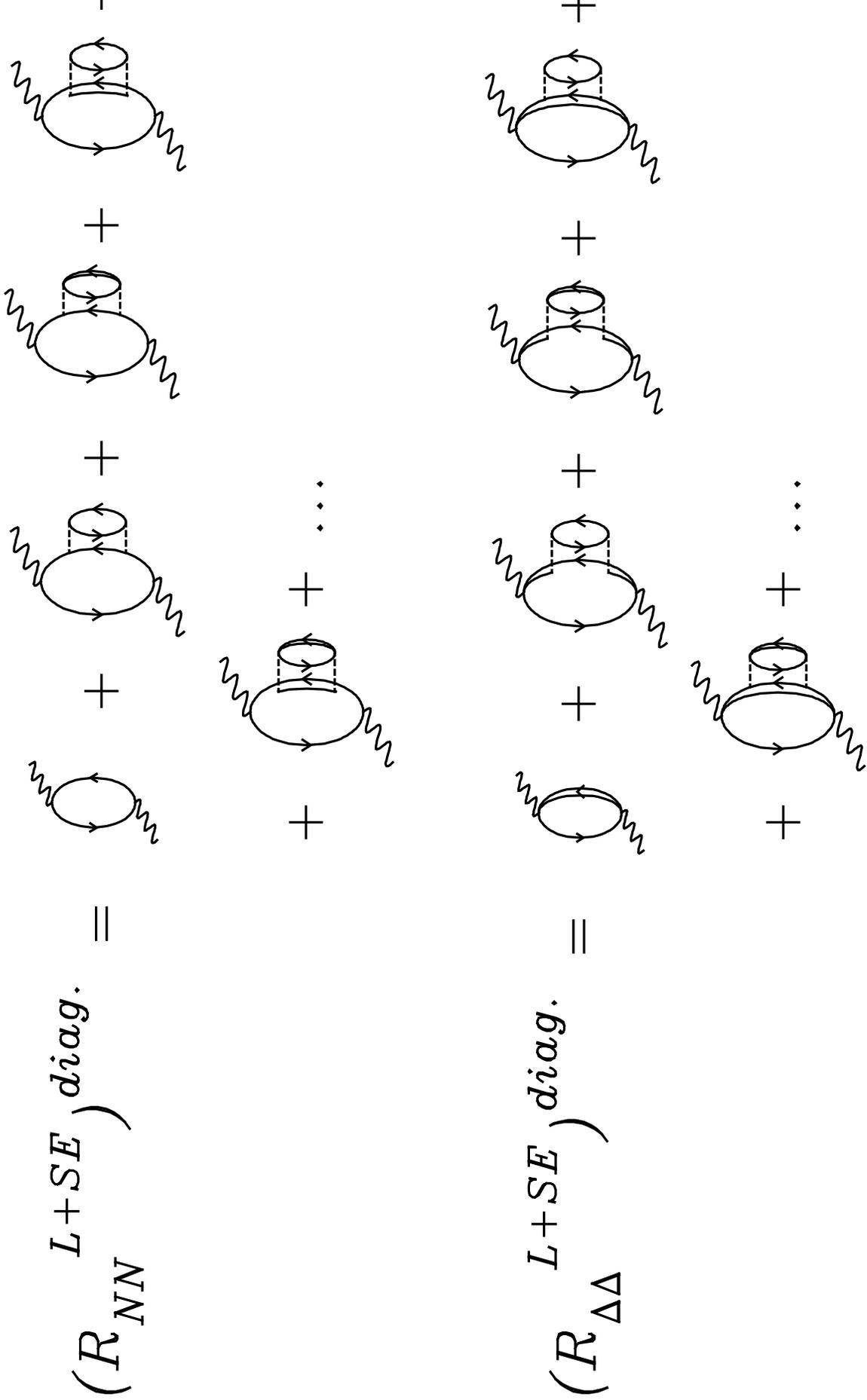
\item [Figure 4:]
Diagonal self energy insertions
from Eq.~(\ref{eq:selfi}).
We show only some of the forward going second order contributions.

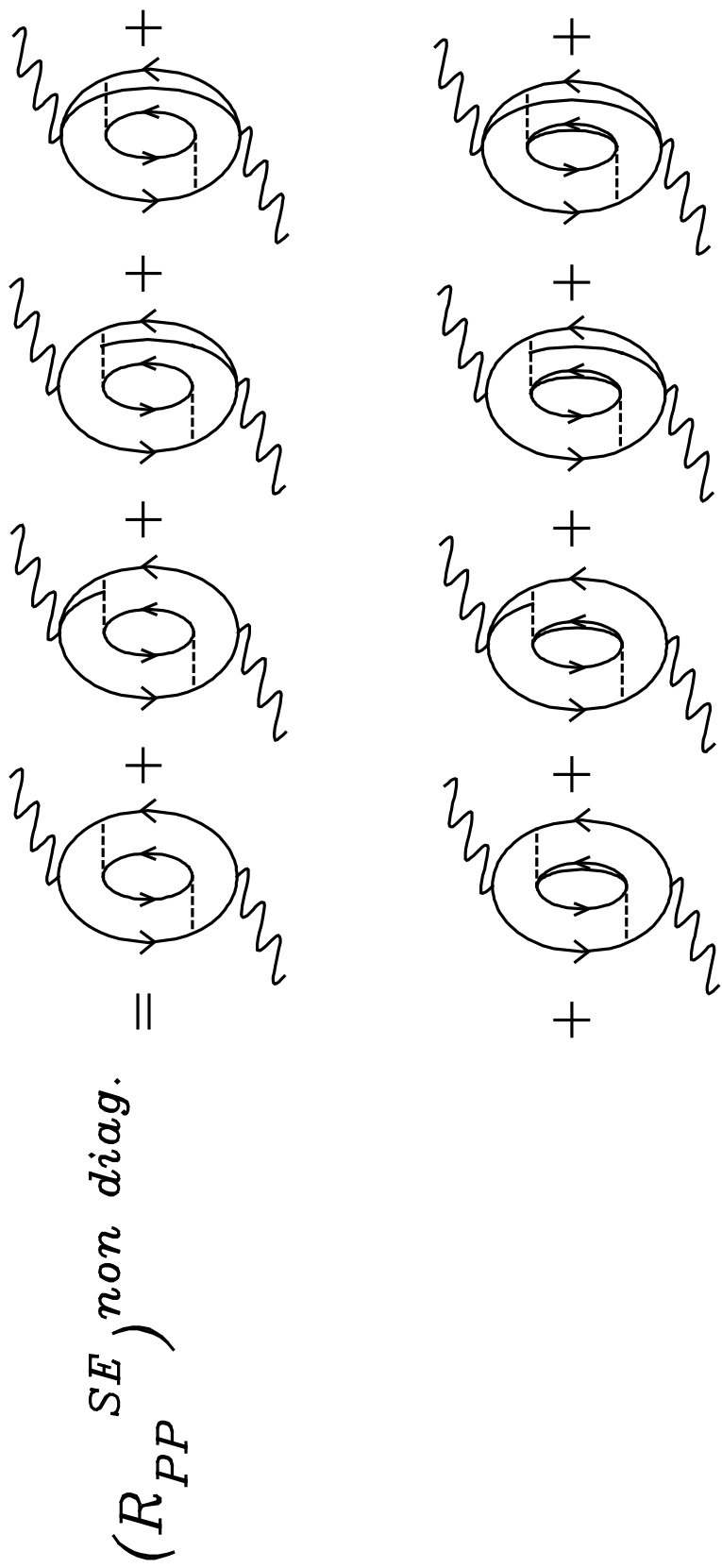
\item [Figure 5:]
Some contributions from the non-diagonal self energy insertions
to the structure function.

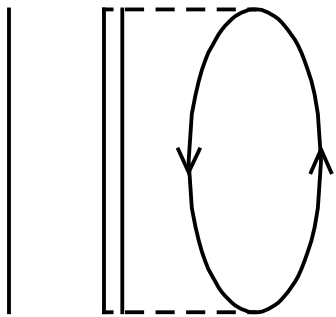
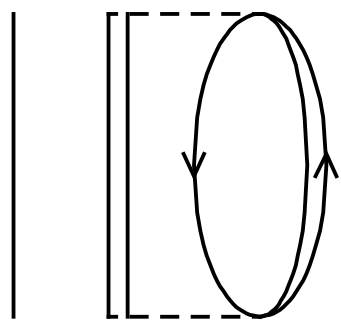
\item [Figure 6:]
Intermediate $Q$-states over which the self energy is sum.
Only direct terms are shown.

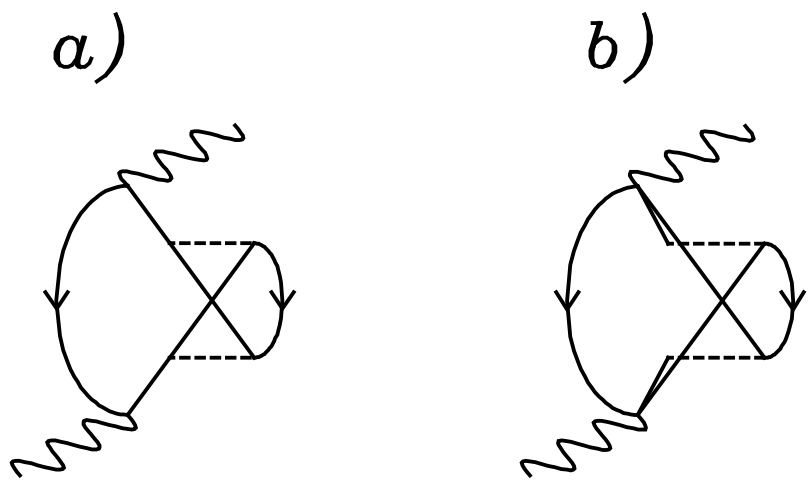
\item [Figure 7:]
Exchange second orders self energy contributions
with intermediate $Q_1$ configuration.

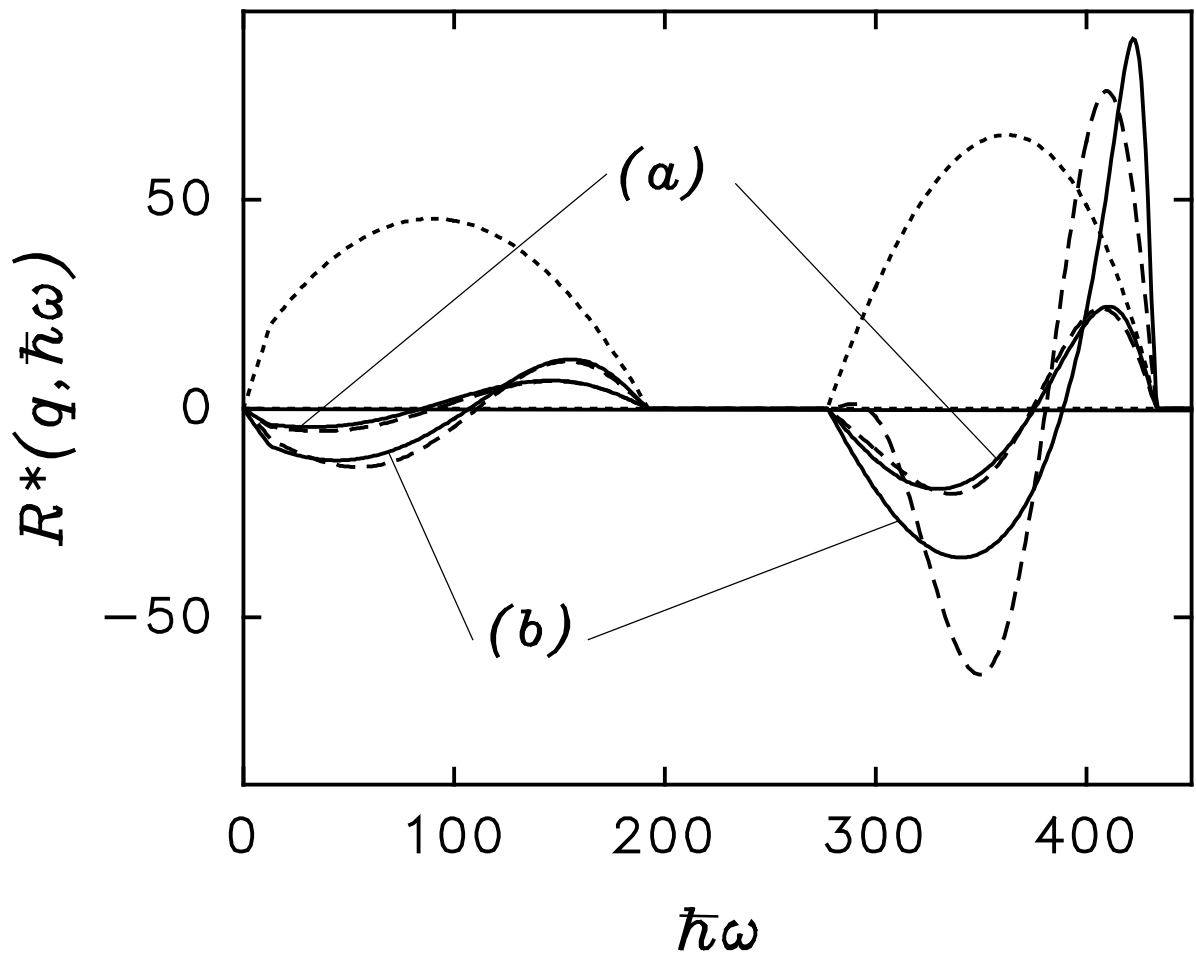
\item [Figure 8:]
Comparison between the RA (continuous lines) and the sum
of terms up to second order (dashed lines) for
$V_a$ (curve label $(a)$) and $V_b$ $(b)$ where
the free structure
function was subtracted from these curves and shown
separately by
dotted lines. Notation $R^* (q, \hbar \omega ) $ indicates
the structure function per unit volume.
The energy is given in $MeV$ and the
structure function in units of $10^{-5} MeV^{-1} fm^{-3}$.

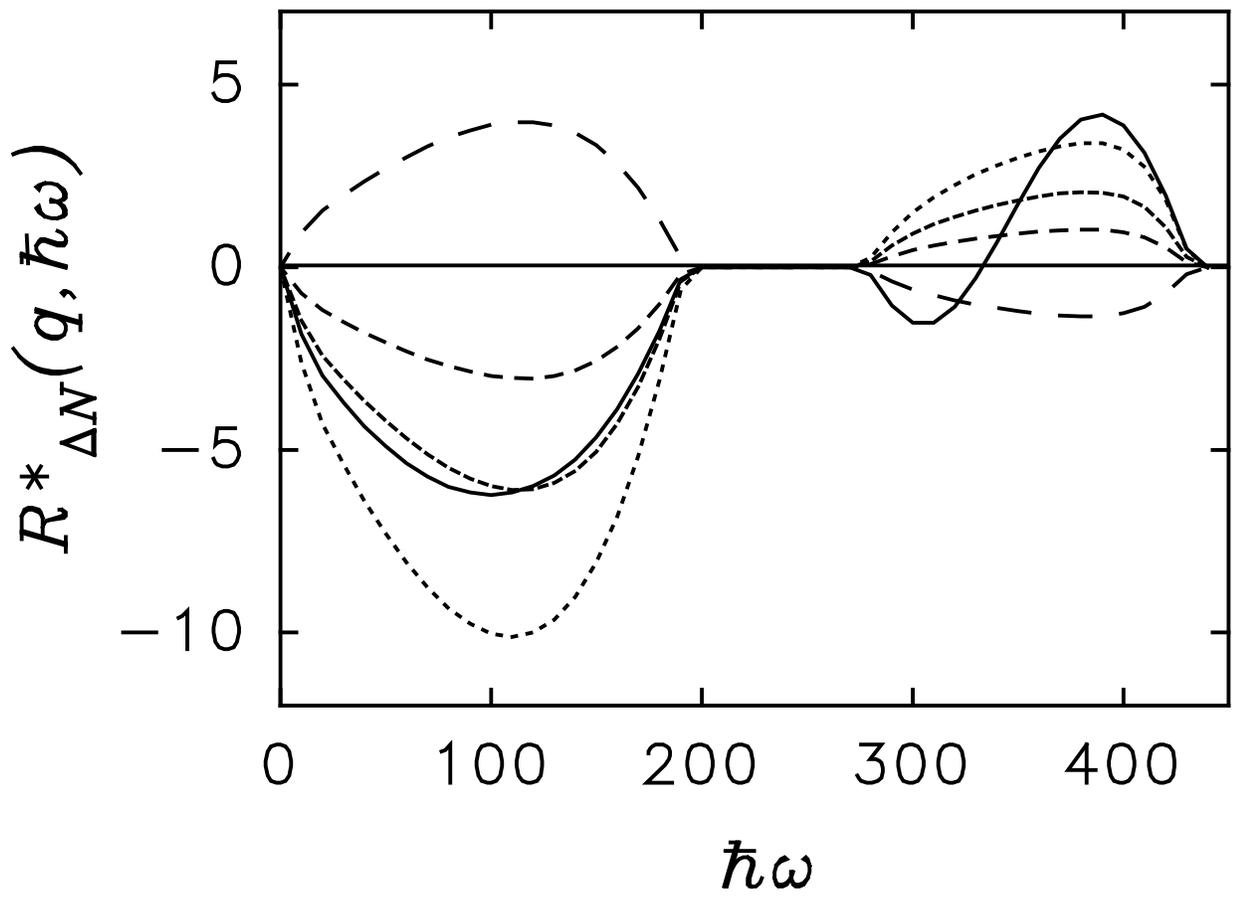
\item [Figure 9:]
$R^*_{\Delta N}$ contribution to the structure function.
The continuous line is our result, $B_{\Delta N}$. Non continuous lines
represent the $\Delta N$ contribution from the RA. Long-dashed
line is for $g'_{\Delta N}=0.0$, dashed line for $g'_{\Delta N}=0.3$,
short dashed line for $g'_{\Delta N}=0.43$ and dotted line
for $g'_{\Delta N}=0.6$. Units are the same as in Fig. 8.

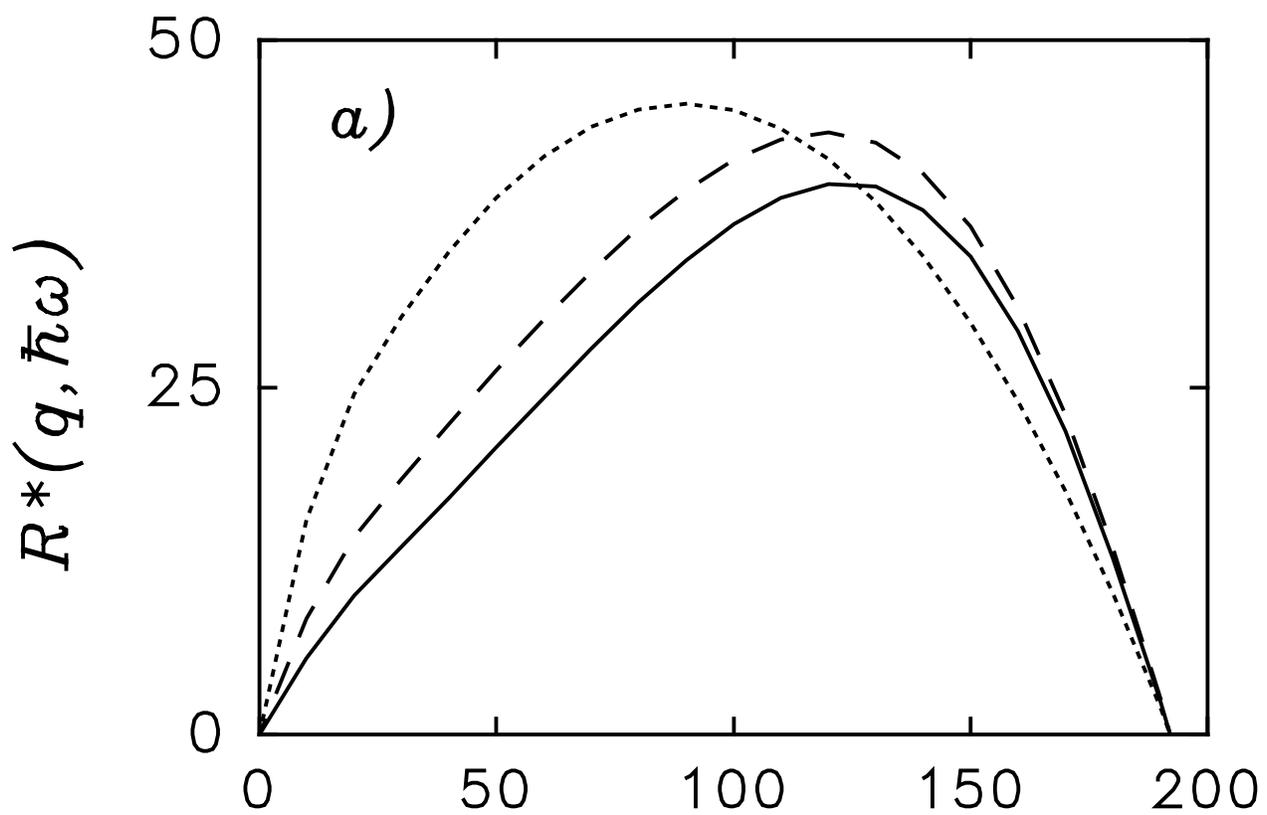
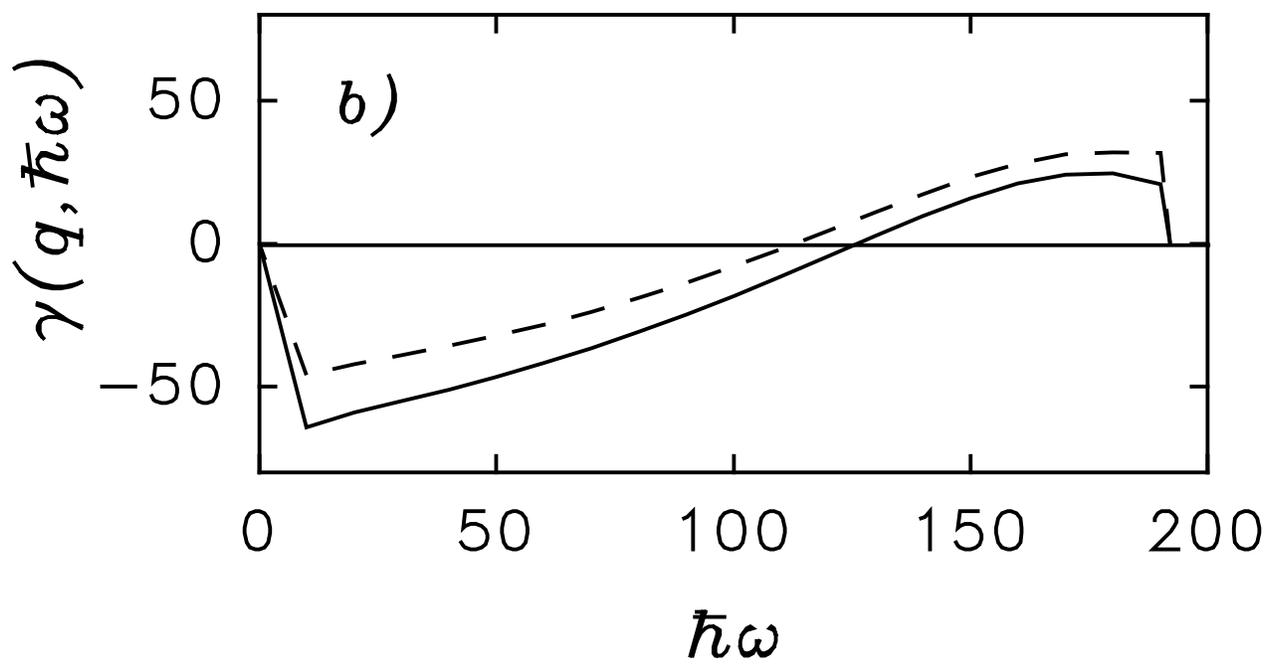
\item [Figure 10:]
RPA result for the structure function. In a), the continuous
line is our result for the RPA with the delta. Dashed line
is the RPA without the delta, while the dotted line
is the free response. In b), we plotted $\gamma$ as defined in
Eq.~(\ref{eq:gamrp}); continuous and dashed lines has the same
meaning as in a).

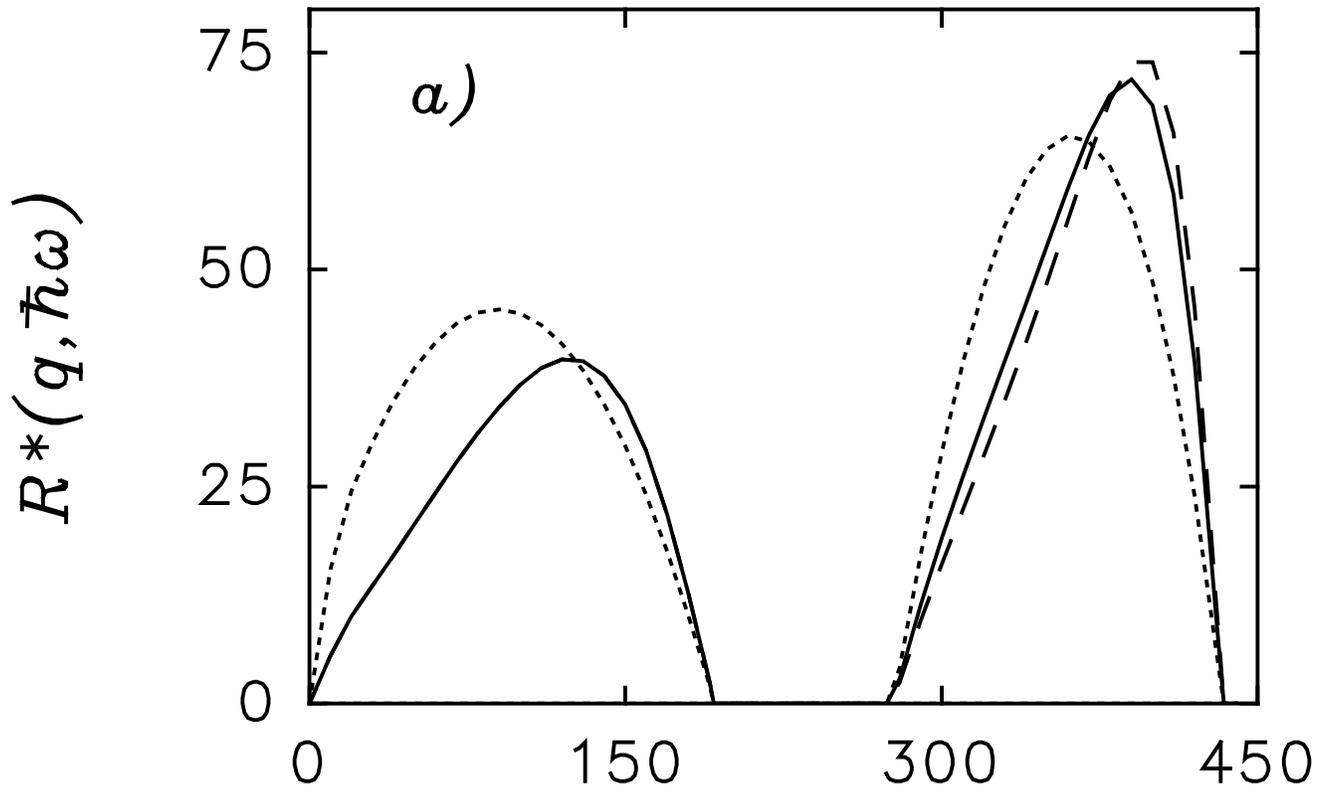
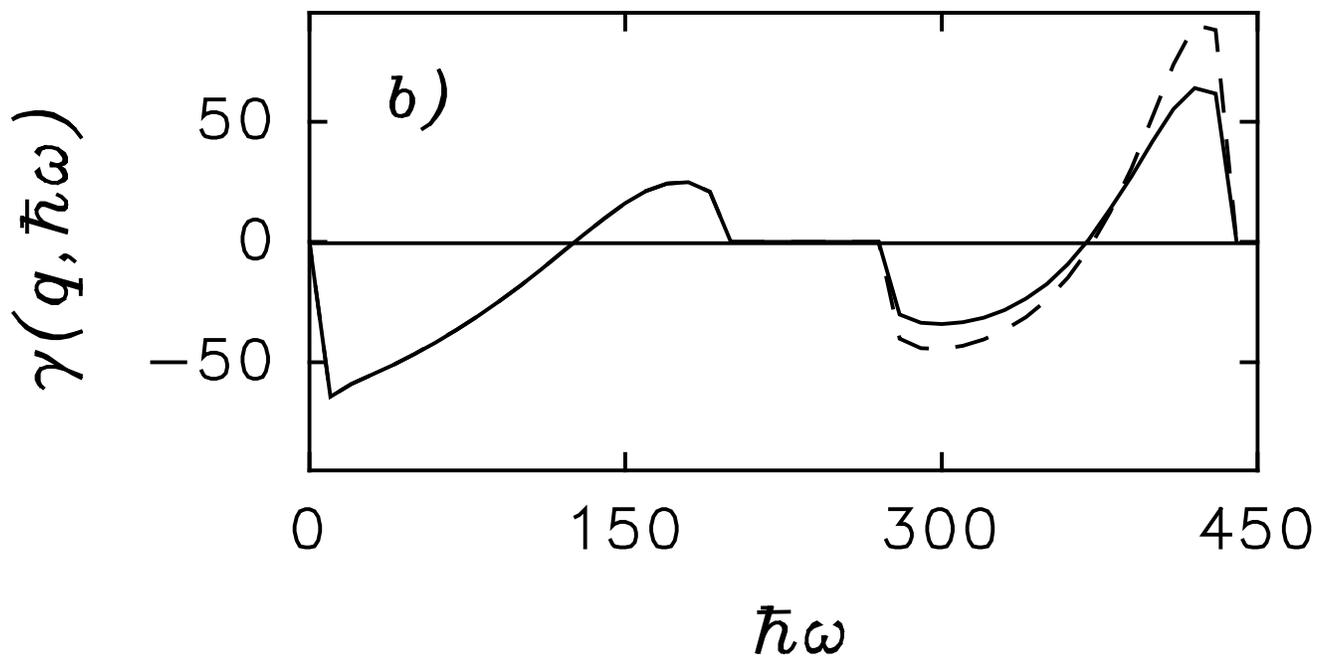
\item [Figure 11:]
RPA result with the delta for two different values of
$g'_{\Delta \Delta}$. In a), continuous line
is for $g'_{\Delta \Delta}=0.4$
and for dashed line $g'_{\Delta \Delta}=0.6$.
Dotted line is the free structure function.
Figure b) has the same meaning as Fig.~10~b.

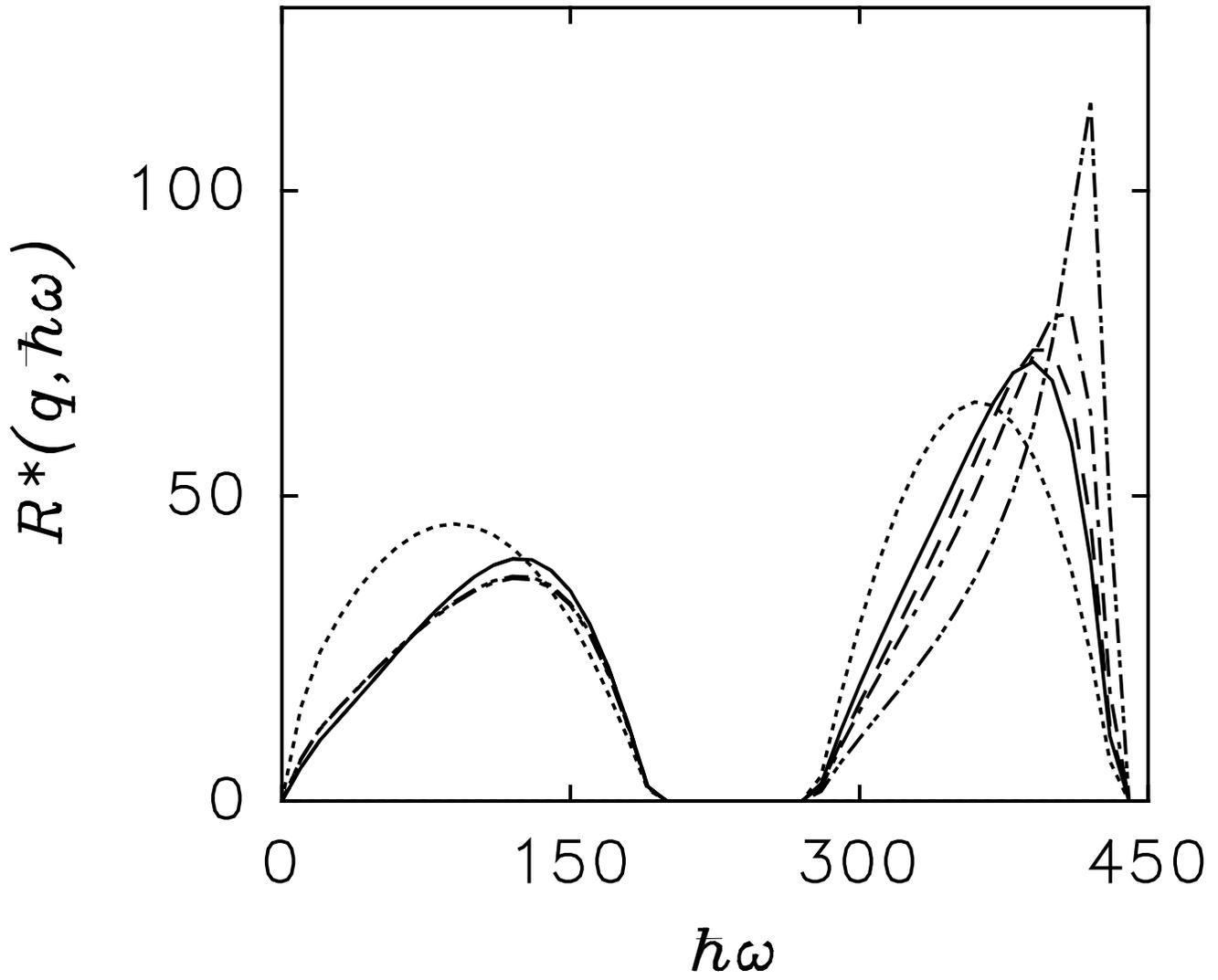
\item [Figure 12:]
Comparison between the RPA and the RA approximations. Continuous
lines is the RPA result for $g'_{\Delta \Delta}=0.4$, dashed
lines is the RPA result for $g'_{\Delta \Delta}=0.6$; dotted
dashed line is the RA result for $g'_{\Delta \Delta}=0.4$
while double dotted-dashed line is the RA result for
$g'_{\Delta \Delta}=0.6$. The remainder parameters of the
interaction are the ones of $V_b$.
Dotted line is the free structure function.

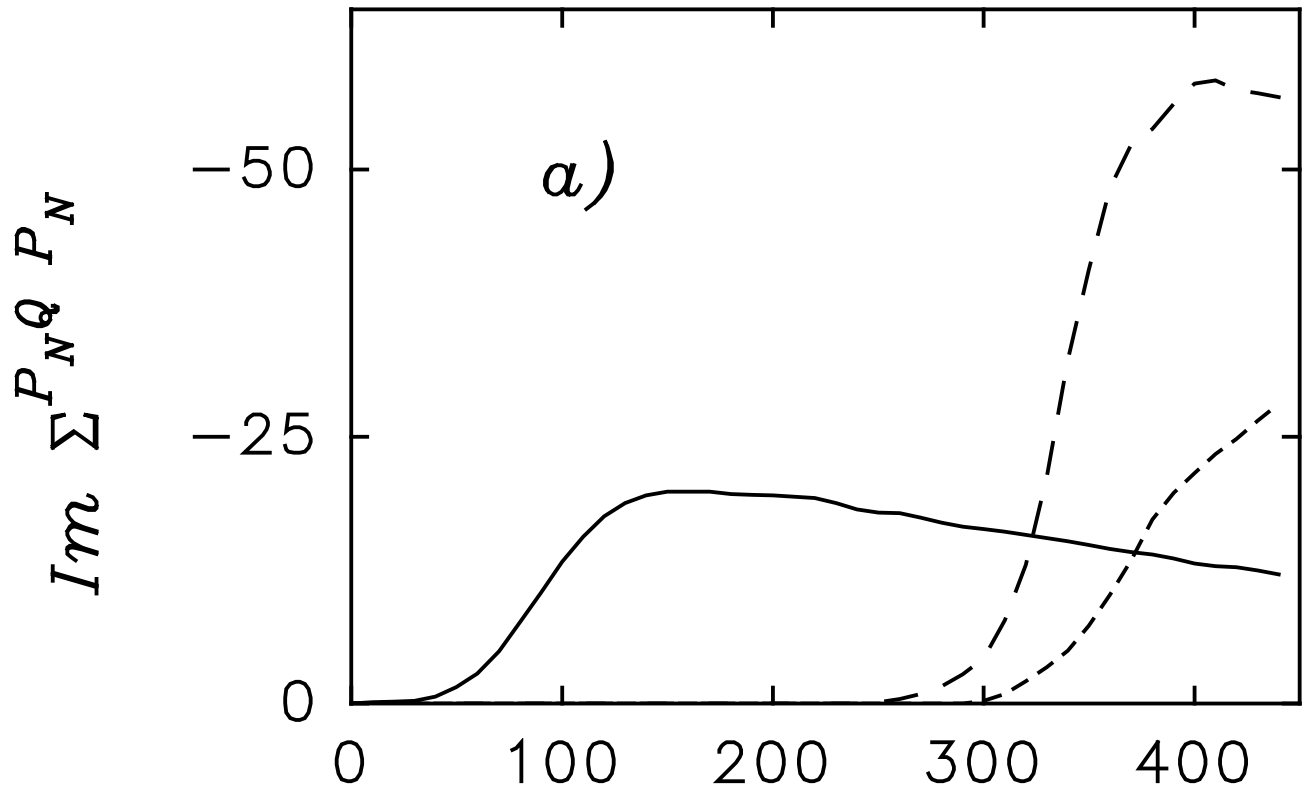
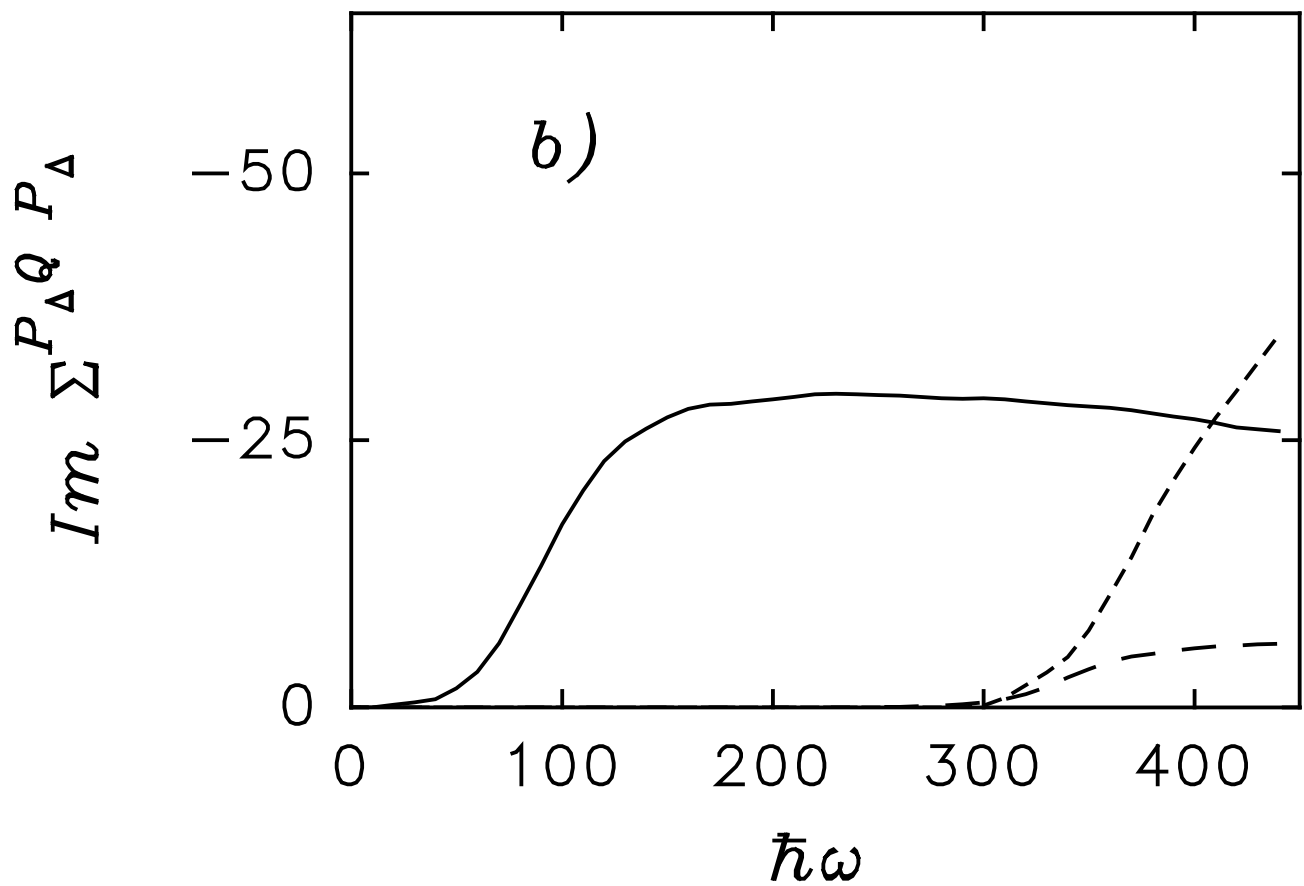
\item [Figure 13:]
Imaginary part of the self energy en MeV. The continuous
line represents the $\Sigma^{P Q_1 P}$ contribution,
the long dashed line is the $\Sigma^{P Q_2 P}$ one while
the short dashed line is the $\Sigma^{P Q_3 P}$ contribution.
In part a) of the figure, $P=P_N$ and in part b) $P=P_{\Delta}$.
The interaction employed was $V_c$.

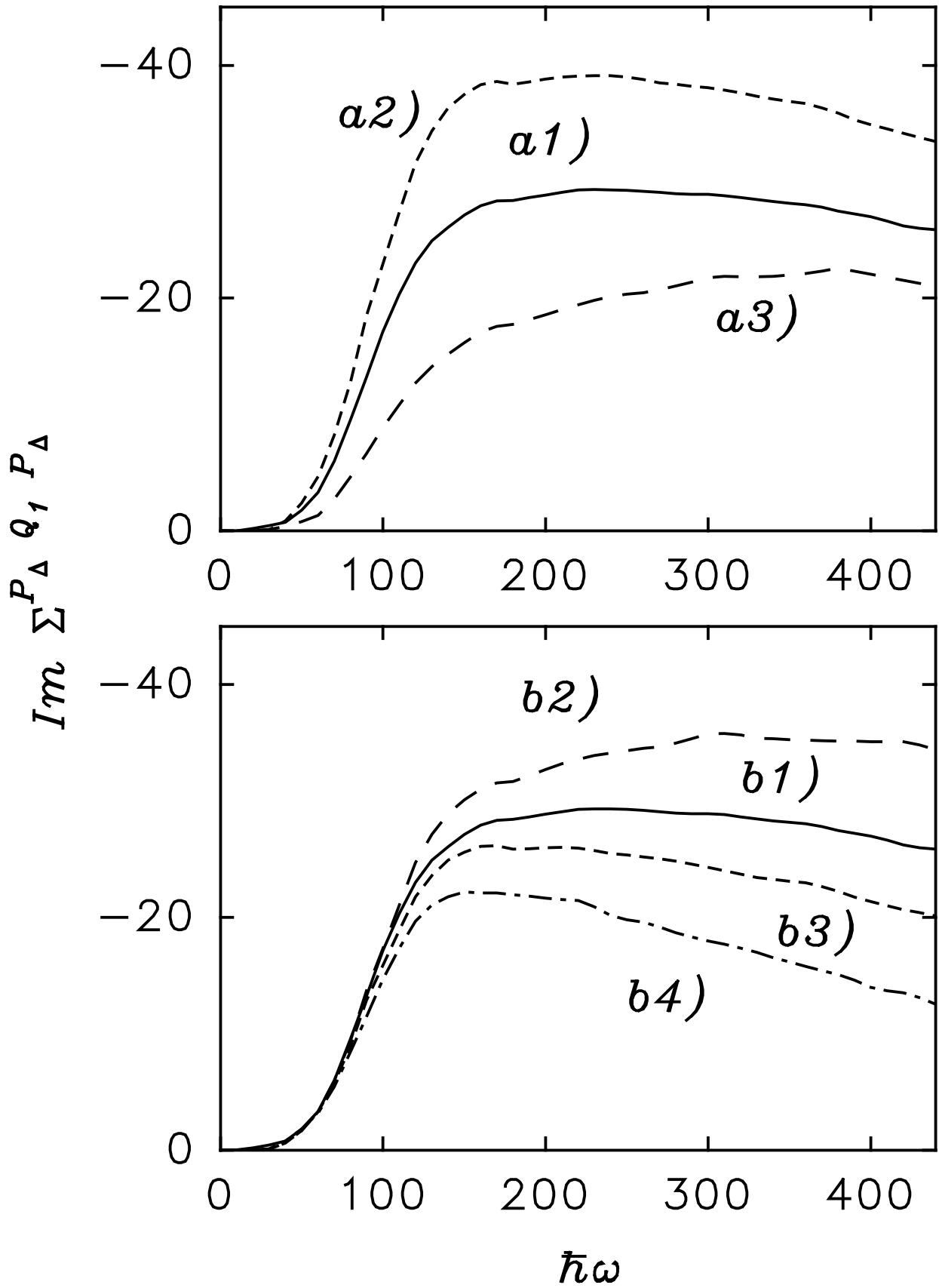
\item [Figure 14:]
Imaginary part of the self energy
$\Sigma^{P_{\Delta} Q_1 P_{\Delta}}$ en MeV.
In both parts, $a1)$ and $b1)$ is our result for $g'_{\Delta N}=0.6$,
$\Lambda_{\pi \Delta N} = \Lambda_{\rho \Delta N}$ = 1000 MeV/c.
While all other parameters are the ones of $V_c$.
All $ai)$ has the same cut offs while $g'_{\Delta N}$ is changed:
0.6 for $a2)$ and 0.4 for $a3)$. And also, all $bi)$ curves has
the same $g'_{\Delta N}$ value and different cut offs:
$\Lambda_{\pi \Delta N} = \Lambda_{\rho \Delta N}$ = 850 MeV/c for $b2)$,
$\Lambda_{\pi \Delta N} = \Lambda_{\rho \Delta N}$ = 1200 MeV/c for $b3)$
and for $b4)$
$\Lambda_{\pi \Delta N}$ = 850 MeV/c and
$\Lambda_{\rho \Delta N}$ = 1200 MeV/c.

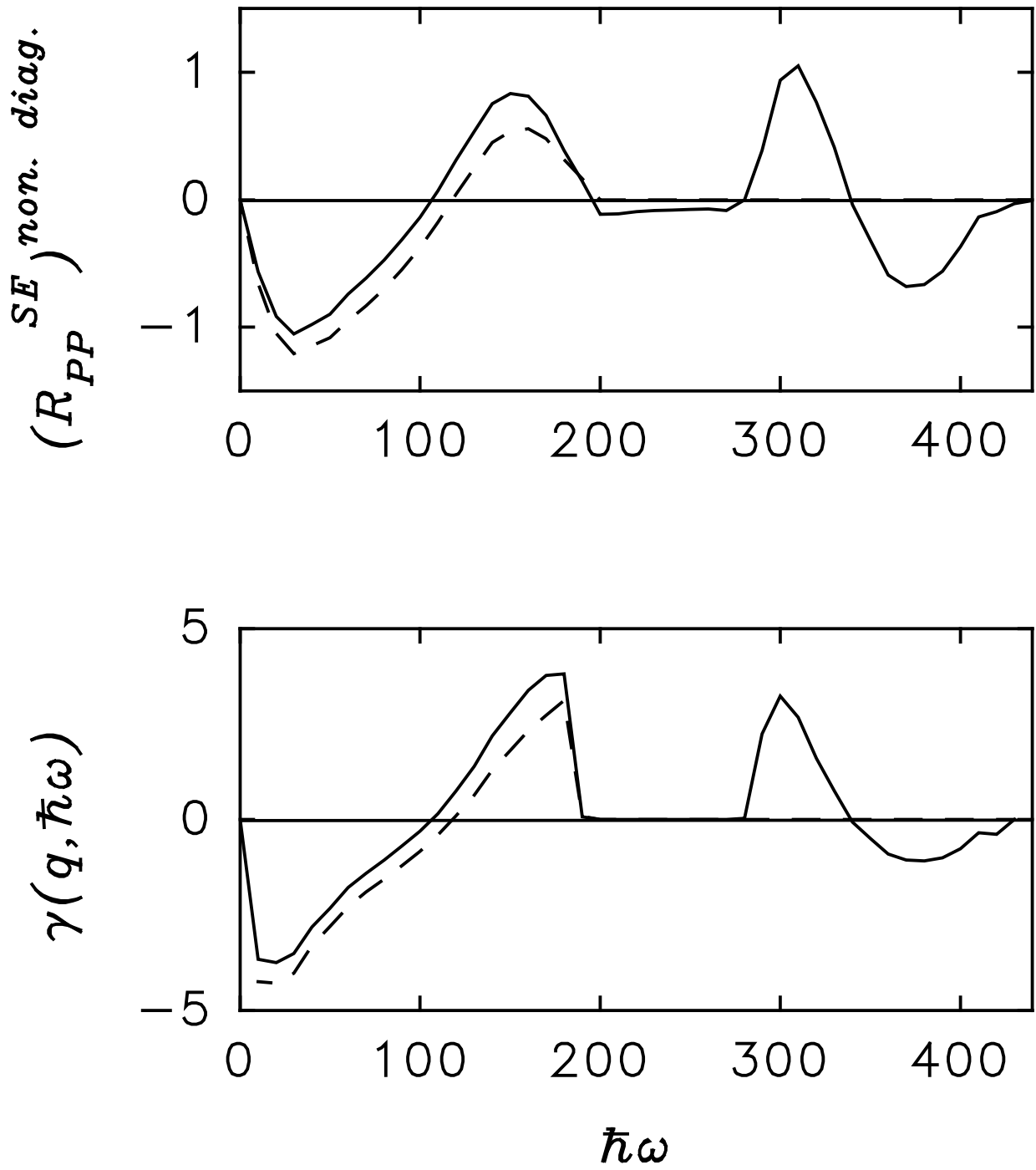
\item [Figure 15:]
Non diagonal self energy contributions to the structure function.
Dashed line is the sum of graphs without the delta and the
continuous line includes it. Also we show $\gamma$ as in
Eq.~(\ref{eq:gamrp}) but where the RPA response was replaced by
the non diagonal self energy contribution.

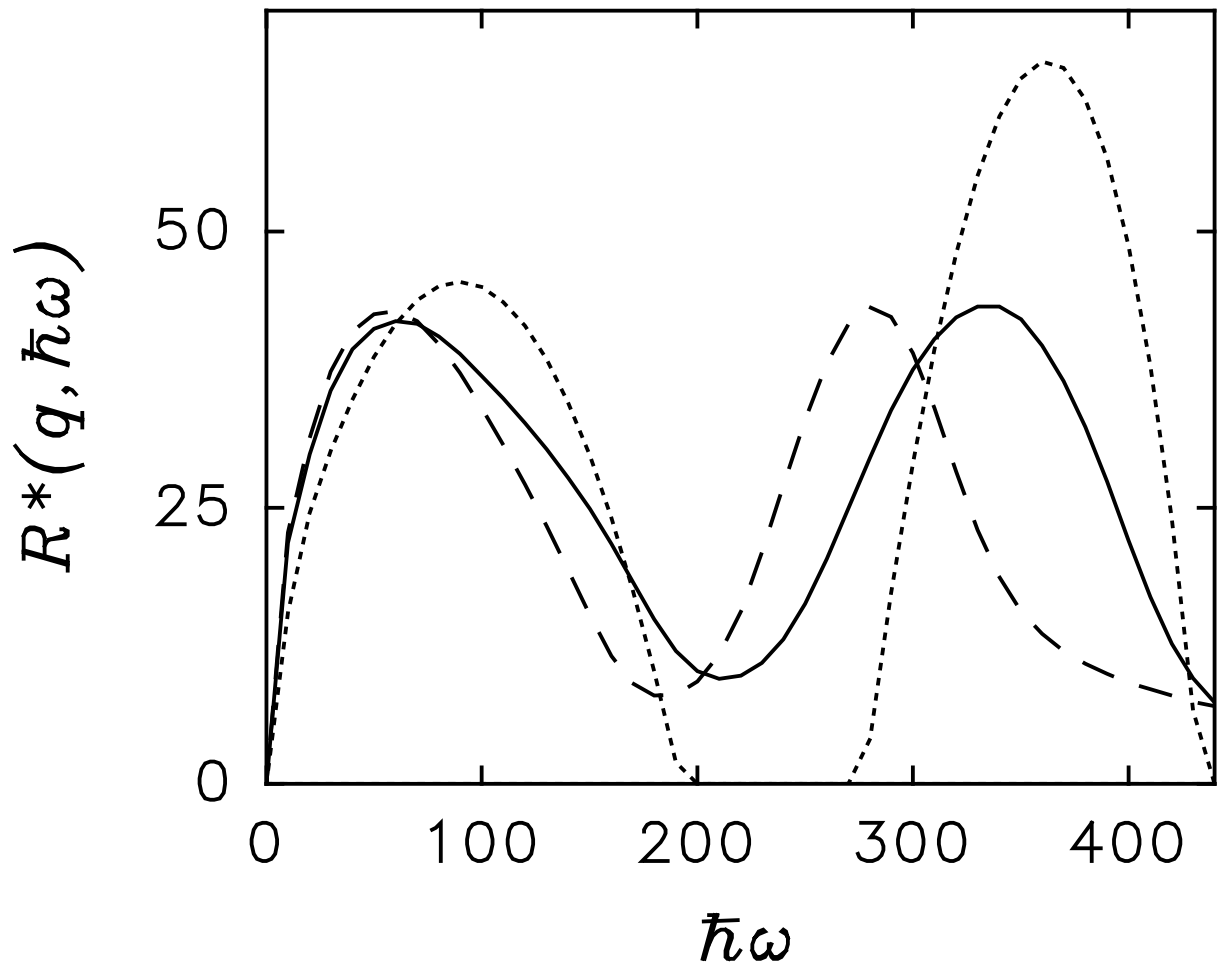
\item [Figure 16:]
Final result for the structure function with
self energies insertions. The dotted line is the
free response. The continuous line is the response function
per unit volume with the $\Sigma^{P Q_1 P}$ self energy insertion.
The dashed line includes all self energy insertions.
Dotted line is the free structure function.

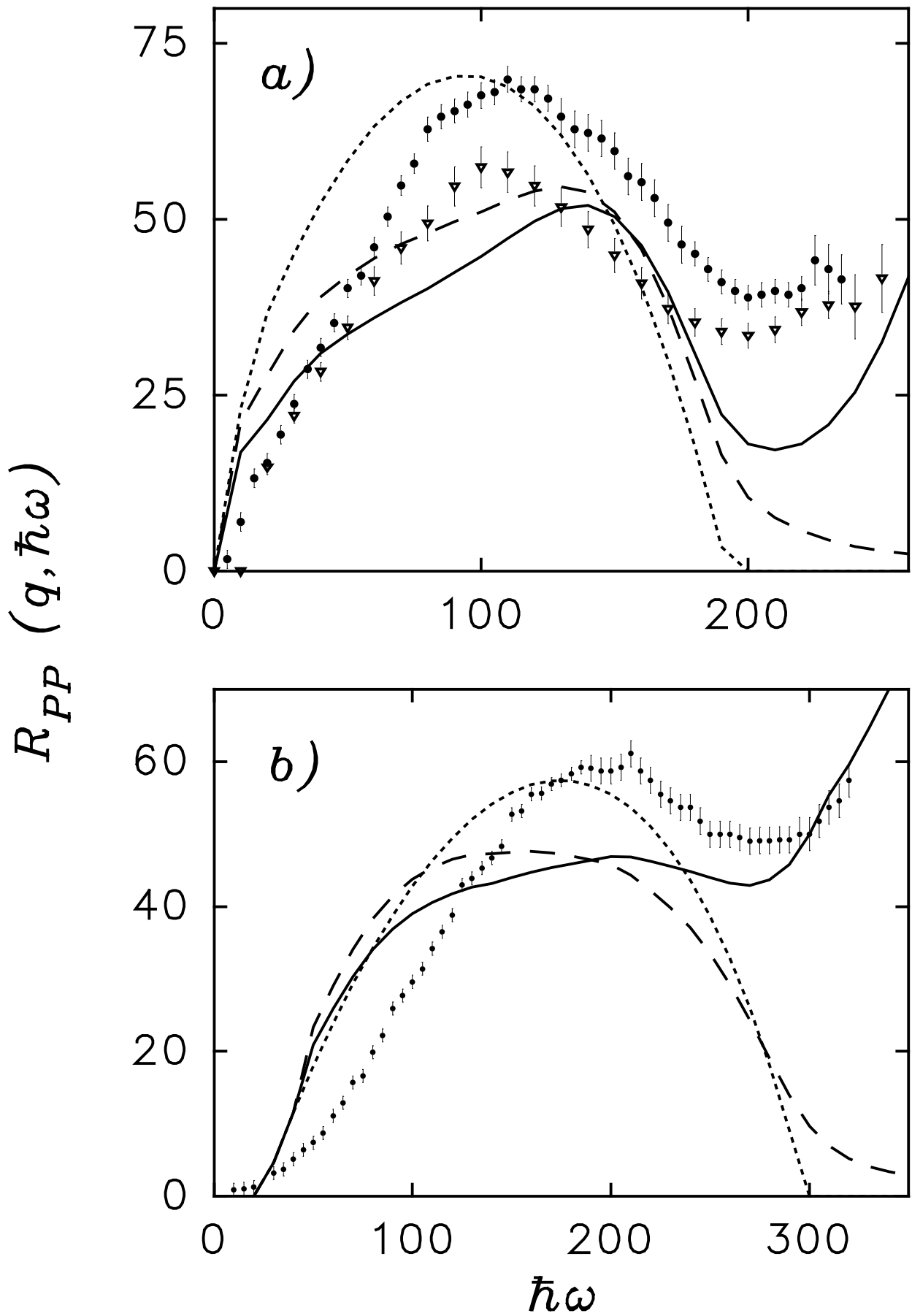
\item [Figure 17:]
Transverse response function for $^{40}Ca$ at momentum transfer
$q=410$ MeV/c for $a)$ and $q=550$ MeV/c for $b)$,
in units of $10^{-3} MeV^{-1}$. Dotted lines
is the free response. Continuous line is our result
from Eq.~(\ref{eq:rpptot}) for the $V_c$
interaction. Dashed line is our result where the delta was excluded.
Data was taken from
\cite{kn:me85} (circles) and
\cite{kn:wi95} (triangles).

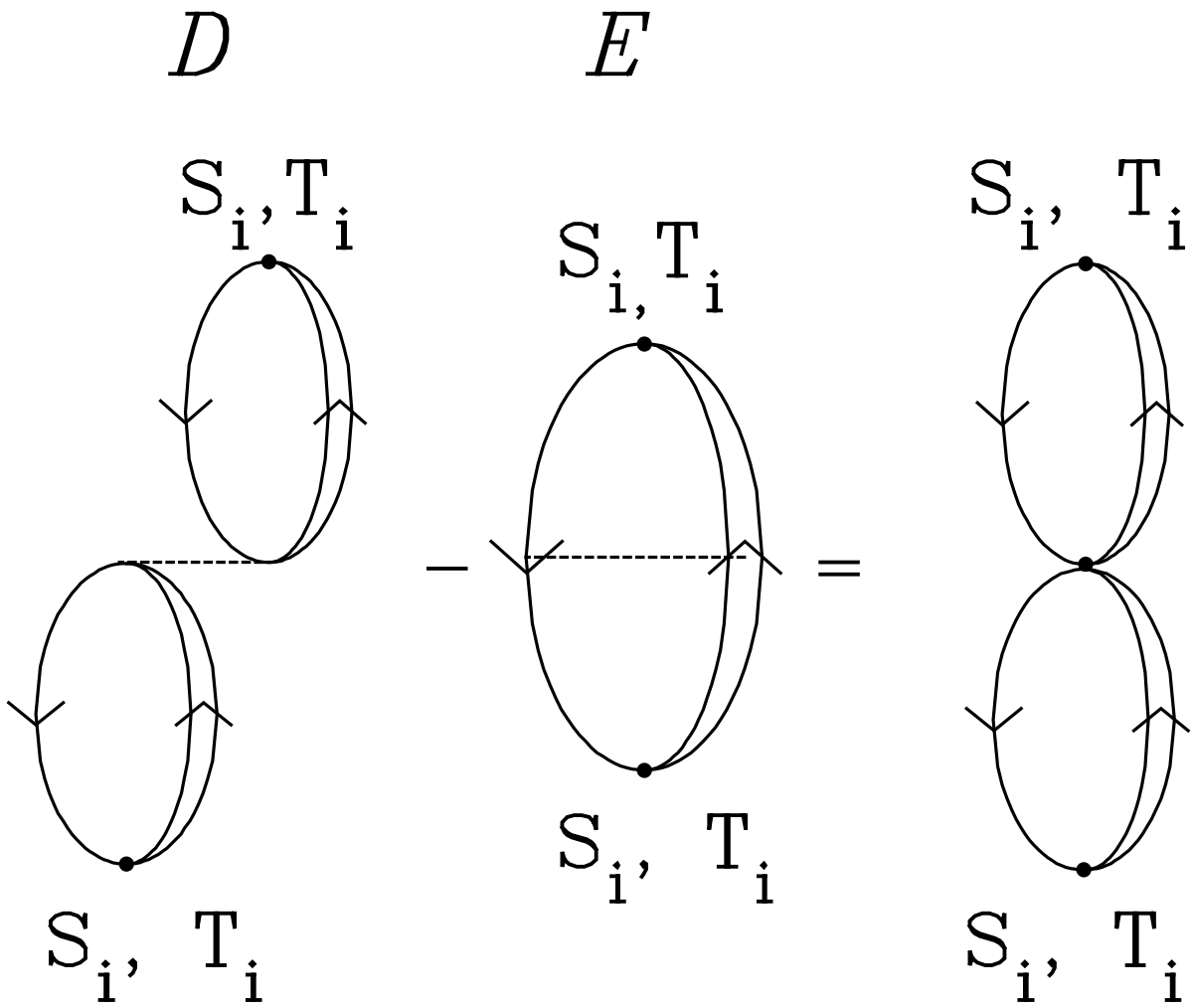
\item [Figure 18:]
Direct plus exchange graphs needed to evaluate
$g'_{1 \, \Delta \Delta}$.
\end{description}
\vfill
\eject

\newpage

\begin{center}

\begin{tabular}{cccccc}       \hline\hline
$V$ & & $g'_{NN}$  & $g_{NN}$ & $g'_{\Delta N}$ & $g'_{\Delta \Delta}$ \\ \hline
$V_a$	   & & 0.5 & 0.0  & 0.0  & 0.3	  \\
$V_b$	   & & 0.7 & 0.3  & 0.0  & 0.5	  \\
$V_c$	   & & 0.7 & 0.3  & 0.6  & 0.4	  \\   \hline\hline \\
\end{tabular}

\end{center}

\vspace{10. mm}

Table I: Values of the Landau Migdal parameters
entering in the different
interactions employed in the text.

\newpage

\begin{center}

\begin{tabular}{cccc}	    \hline\hline
\multicolumn{4}{c} {$B_{1 \, NN}$}   \\ \hline
$V$ & $\hbar \omega$ & $dir.$	& $exc.$  \\ \hline
$V_a$	  &   ~40.	   &  ~-8.17  &  ~~2.36 \\
	  &   150.	   &  ~~3.38  &  ~-1.12 \\
$V_b$	  &   ~40.	   &  -18.90  &  ~~5.68 \\
	  &   150.	   &  ~~7.84  &  ~-2.48 \\   \hline \\
\multicolumn{4}{c} {$B_{1 \, \Delta \Delta}$}	\\ \hline
$V$ & $\hbar \omega$ & $dir.$	& $exc.$  \\ \hline
$V_a$	  &   320.	   &  -21.45  &  ~~5.51 \\
	  &   400.	   &  ~20.83  &  ~-5.35 \\
$V_b$	  &   320.	   &  -41.21  &  ~67.25 \\
	  &   400.	   &  ~40.04  &  -58.91 \\   \hline\hline
\end{tabular}

\end{center}

\vspace{10. mm}

Table II: Direct and exchange first order contributions from the
RPA series to the structure function
at momentum transfer $q=410 \; MeV/c$.
The residual interaction is taken from Table I.
We have considered the $NN$ and $\Delta \Delta$ channels.
The energy is given in $MeV$ and the
structure function in units of $10^{-5} MeV^{-1} fm^{-3}$.
The values of the energy where taken closer
to the points where each particular response has its extreme.

\newpage

\begin{center}

\begin{tabular}{ccccc}	     \hline\hline
\multicolumn{5}{c} {$B_{1 \, \Delta N}$}   \\ \hline
$V$  & $\hbar \omega$ & $dir.$	 & $exc.$  & $exc \, (w.t.) $  \\ \hline
$V_a$	    &	100.	     &	 ~4.37	&   -8.30 &  -1.99 \\
	    &	340.	     &	 -1.55	&   ~3.01 &  ~0.79 \\	\hline \\
\multicolumn{5}{c} {$B_{2 \, \Delta N}$}   \\ \hline
$V$ & $\hbar \omega$ & $dir.$	& $exc.$ & $exc \, (w.t.) $  \\ \hline
$V_a$	   &   ~50.	    &	-1.22  &   ~1.69  & ~0.57   \\
	   &   100.	    &	-0.78  &   ~2.05  & ~0.54   \\
	   &   320.	    &	~1.25  &   -3.90  & -0.99   \\
	   &   400.	    &	-0.75  &   ~2.34  & ~0.55   \\	 \hline\hline
\end{tabular}

\end{center}

\vspace{10. mm}

Table III: The same as Table II, but for the $\Delta N$
channel. We have also included second order
contributions $B_{2 \, \Delta N}$. In the last column $exc (w.t.) $,
we have shown the exchange term without the tensor contribution
to the residual interaction.

\newpage

\begin{center}

\begin{tabular}{ccc}	   \hline\hline
\multicolumn{3}{c} {$Im \, \Sigma^{P_N Q_1 P_N}$}   \\ \hline
 $\hbar \omega$ & $dir.$   & $exc.$  \\ \hline
  100.	       &  -13.31  &    0.20 \\
  200.	       &  -19.46  &    1.12 \\
  300.	       &  -16.34  &    0.83 \\
  400.	       &  -13.12  &    0.23 \\	 \hline \\
\multicolumn{3}{c} {$Im \, \Sigma^{P_{\Delta} Q_1 P_{\Delta}}$}   \\ \hline
 $\hbar \omega$ & $dir.$   & $exc.$  \\ \hline
  100.	       &  -17.14  &    0.80 \\
  200.	       &  -28.85  &    3.08 \\
  300.	       &  -28.92  &    1.89 \\
  400.	       &  -26.98  &    0.63 \\	 \hline\hline
\end{tabular}

\end{center}

\vspace{10. mm}

Table IV: Direct and exchange self energy contributions in units
of MeV for several energies.

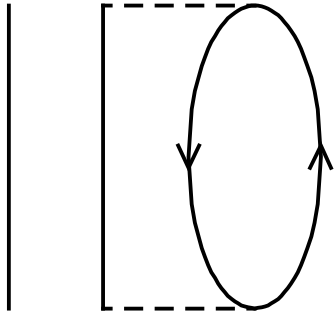
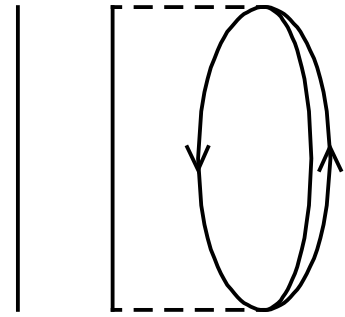
\end{document}